# Correlated Plasmonic Excitation in Twisted Nematic Plasmonic Superlattices

Rui Huang[‡,1], Jordan Austin-Frank Wilson[‡,1], Anders Dollard[1], Nicholas C Fisher[1], Zixian Fang[1], John T. Fourkas[1,2,3], Zhiwei Li[*,1]

1Department of Chemistry and Biochemistry, University of Maryland, College Park, Maryland 20714, United States

2Institute for Physical Science and Technology, University of Maryland, College Park, Maryland 20714, United States

3Maryland Quantum Materials Center, University of Maryland, College Park, Maryland 20714, United States

Email: zlinano1@umd.edu

‡ These authors contributed equally.

**Abstract**: Superlattices with twisted configurations, such as moiré lattices, have recently been extensively exploited for their unique electronic, magnetic, and optical properties. One remarkable feature of nanoscale twisted superlattices is the distinct lattice symmetries and the continuous phase transitions between periodic or aperiodic phases, representing a unique opportunity to study many emerging physical phenomena. Here, we report a correlated light–matter interaction between the collective polarization effect of nematic plasmonic superstructures and the plasmonic excitation of individual constituent nanorods in reconfigurable twisted plasmonic superlattices. Using hybrid $Fe_3O_4$/Au nanorods as building blocks, we assembled plasmonic nematic liquid crystals with unidirectionally aligned nanorods, which could be further assembled into moiré plasmonic lattices through a vertical stacking assembly method. A twist-angle-dependent plasmonic excitation is recognized in the twisted bilayer of two plasmonic superlattices, featuring enhanced transverse and longitudinal plasmonic excitation at a twisting angle of 0° and 90°, respectively. Such correlated plasmonic excitation in twisted plasmonic superstructures is induced by the correlation between the collective polarization effect of the liquid-crystal phases and the anisotropic plasmonic excitation of individual nanorods. The magnetic orientation control allows for precise alignment of hybrid $Fe_3O_4$/Au nanorods in polymer substrates and enables the coding of nematic domains and plasmonic patterns in each sublattice. The correlated plasmonic excitation and light polarization create reconfigurable photonic moiré superlattices with well-defined domain colors, feature sizes, periodicities, symmetries, and dimensions determined by twist angles and displacements in the twisted plasmonic lattices.

## Introduction

Moiré superlattices comprise two superimposed sublattices with a relative twist angle and/or translational displacement.[1-5] These superstructures are inspired by geometric moiré patterns that were originally introduced for describing perceivable interference patterns of two regular images in mathematics and were later extended to materials science and physics for their unique translational and rotational symmetries in the resulting periodic and aperiodic structures.[6-10] A few famous examples in this regard include twisted bilayer graphene[11-17] and moiré superlattices of many two-dimensional (2D) van der Waals heterostructures,[18-22] which can exhibit remarkable electronic, magnetic, and optical properties. Moiré photonic superlattices have also been reported to have many emerging optical and photonic properties that were previously challenging to realize experimentally.[23-29] Moiré superlattices comprising two square sub-lattices demonstrate unconventional light localization and delocalization based on flat-band physics, producing tunable light patterns in response to twist angles.[1, 30] Collective and hybridized excitation is also possible in twisted bilayers due to optical transitions between quasi-localized states from the moiré superlattices.[31-33] In a carefully prepared study, domain reconstruction in moiré superlattices was reported in twisted $MoSe_2$/$MoSe_2$ bilayers, with mirror symmetry-breaking excitons due to the mismatch of atomic configuration in each domain.[34] Such field localization and excitation are promising in designing metalenses,[35, 36] exotic metasurfaces,[37] collective excitonic phases,[38] and many smart optical materials.[39] Understanding how these unique light states and polarization impact the nanoscale light–matter interactions in nanostructured materials is the key to realizing these compelling applications of moiré metamaterials.

Here, we report correlated excitation of a localized surface plasmon resonance (LSPR) in moiré plasmonic superlattices by controlling the twist angles and translational displacement in a stacked bilayer of nematic plasmonic superlattices. Hybrid $Fe_3O_4$/Au nanorods are used as building blocks, and the coupled magnetic and plasmonic anisotropy enables unidirectional alignment of these nanorods into nematic plasmonic liquid crystals and further encoding into patterned sublattices. By stacking two nematic plasmonic superlattices, we created moiré plasmonic superlattices with long-range twisted structural orders. In such twisted plasmonic bilayers, we observed a correlated plasmonic excitation, featuring twist-angle-dependent excitation strength of transverse and longitudinal modes and color changes in response to mechanical twisting. Our systematic studies on the spectral properties and light polarization states lead to the conclusion that the correlated plasmonic excitation is induced by the correlation of collective light polarization of the nematic phases and the anisotropic plasmonic excitation of constitutive Au nanorods. Using the correlated LSPR, we demonstrate highly reconfigurable moiré plasmonic superlattices with theoretically unlimited and experimentally defined symmetry, periodicity, feature size, domain color, and dimensions.

**Correlated plasmonic excitation in twisted nematic plasmonic superlattices**

The building blocks for creating the twisted superlattices are hybrid $Fe_3O_4$/Au@RF nanorods, which are synthesized using a templating method.[40-42] The building blocks comprise one magnetic nanorod and one plasmonic Au nanorod that are aligned in parallel within a resorcinol-formaldehyde (RF) polymer shell (inset in **Figure 1a**). The morphology of hybrid nanorods with an Au-nanorod aspect ratio of ~1.8 is shown in the TEM image in **Figure 1a**. The extinction spectrum has a maximum for the longitudinal mode at λL = 620 nm (**Figure S1**). Relying on the coupled magnetic and plasmonic anisotropy, the hybrid nanorods can be unidirectionally aligned in a solid film using a magnetic field-assisted lithography method.[43-45] During the photopolymerization of polyacrylamide (PAM), a magnetic field was applied so that the hybrid nanorods were parallel to the field.[46, 47] Such a nematic phase of aligned Au nanorods creates an orientation-dependent plasmonic excitation of the resulting plasmonic films, leading to peak absorption changes under linearly polarized light.[48-51] When the rod orientation changes from parallel (0$^o$) to perpendicular (90$^o$) to the polarization, the absorption of longitudinal and transverse peaks decreases and increases, respectively (**Figure 1b**). At 0$^o$, the parallel light polarization only excites free-electron oscillations along the nanorods' long axes, creating selective longitudinal excitation and maximum peak intensity at λL. Selective transverse excitation is realized at 90$^o$, for which the nanorods are perpendicular to the light polarization. The selective excitation enables color switching in the plasmonic superlattices by controlling the light polarization. Applying a parallel or perpendicular polarizer causes a color change to green or red, respectively; these colors are complementary to $\lambda_L$ and to $\lambda_T$ = 520 nm, the maximum of the transverse peak (**Figure 1c**).

Stacking two identical plasmonic films induces a correlated excitation in the twisted bilayer, making the peak absorbance dependent on the twisting angle under excitation of unpolarized light. To describe the stacking effect quantitatively, the twist angle is defined as the angle between the nanorod alignment in the two layers of the bilayer (inset in **Figure 1d**). When the stacking angle increases to 90$^o$ clockwise, the absorption in the longitudinal and transverse peaks increases and decreases under excitation of unpolarized light, respectively (**Figure 1d**). Further theoretical analysis (see supplementary information for a detailed derivation) indicates that the longitudinal plasmonic extinction at a 90$^o$ twist approaches ~80% of the theoretical maximum compared with the extinction of a single plasmonic film illuminated with linearly polarized light (**Figure 1e**). This theoretical consideration suggests that the extinction by the longitudinal and transverse modes is enhanced at large (between 45$^o$ and 90$^o$) and small twist angles (between 0$^o$ and 45$^o$), respectively. The enhanced plasmonic extinction induces defined color changes at different twist angles. As demonstrated in **Figure 1f**, the enhanced transverse extinction produces a complementary red color at 0°, whereas the enhanced longitudinal extinction makes the stacked areas appear blue at 90°. Leveraging the flexible magnetic assembly, we can create plasmonic patterned films that feature orthogonal nanorod alignment in neighboring domains. Shown in **Figure S2** is a sample in which two different patterns were created by magnetically aligning the hybrid nanorods in the horizontal and vertical directions in blue and red domains, respectively. If two such films are assembled into twisted superlattices, complicated, twist-angle-dependent moiré patterns appear solely in overlapped regions, and there are no perceived domains in the non-overlapping regions under unpolarized illumination in either AA stacking (**Figures 1g** and **h**) or AB stacking (**Figure 1i**). In non-overlapping regions, the plasmonic extinction in domains of vertical and horizontal nanorod alignment is the same, making the domain boundaries indistinguishable. In the overlapping regions, the red domains are expected to have enhanced transverse mode extinction, and the blue domains are expected to have enhanced longitudinal mode extinction, based on the complementary color principle.

**Wavelength-dependent polarization states in nematic plasmonic lattices**

To study the twist-angle-dependent plasmonic excitation, we measured the polarization states of transmitted light by using rotating quarter-wave-plate Stokes polarimetry (**Figure 2a**). Light of a specific wavelength is incident on the samples, and the light intensity was measured by changing the rotation angle of the quarter-wave plate from 0° to 180°. **Figure 2b** plots the transmitted light intensity of vertically aligned and random hybrid nanorod samples at different rotation angles of the plate, demonstrating periodic intensity changes in the aligned sample. Using the home-built polarimeter, we can derive the Stokes parameters ($S_0$, $S_1$, $S_2$, $S_3$) from intensity modulation measured as the plate rotates. The degree of polarization (DOP) of a light polarization state can be calculated by

$$\frac{\sqrt{S_1^2 + S_2^2 + S_3^2}}{S_0}$$

. As illustrated in **Figure 2c**, the polarization state of light can be defined using the polarization ellipse representation, with ψ and χ being the orientation angle and ellipticity angle, respectively. Under unpolarized light, both the magnetic nanorods and hybrid nanorods change the polarization state of the incident light, but through different mechanisms. We measured the DOP of light with and without the nanorod samples and calculated the changes in DOP (ΔDOP) to understand how these samples dictate light polarization states. During the measurement, a horizontal (**Figure S3a**), vertical (**Figure S3b**), and parallel (**Figure S3c**) magnetic field was applied along the y-, z-, and x-axes, respectively, and the light was incident along the x-axis (inset in **Figure 2b**). As shown in **Figure 2d**, the ΔDOP decreases as the wavelength of incident light increases in nematic liquid crystals made of magnetic $Fe_3O_4$@$SiO_2$ nanorods.

The polarization effect of the magnetic liquid crystals demonstrates that the transmitted light is polarized perpendicular to the nanorods' alignment; this phenomenon is induced by the birefringence that results from the orientational order in liquid crystals.[52] When the $Fe_3O_4$/Au hybrid nanorods were used as building blocks, we observed interesting wavelength-dependent DOP changes in the transmission of light. Near λT, the transmitted light is polarized parallel to the nanorod alignment. Near λL, the transmitted polarization is perpendicular to the rod alignment. **Figure 2e** shows that the polarization direction switches from being parallel to being perpendicular to the nanorod alignment when the wavelength of the incident light increases from λT to λL. The effect of nanorod alignment on the polarization state can be directly characterized using the polarization-ellipse representation derived from the Stokes polarimetry. At λL, the polarization of light is always perpendicular to the nanorods' alignment (**Figure 2f**). As shown in **Figure S3d**, at short wavelengths, the polarization is switched to vertical by horizontally aligned nanorods due to light–matter interactions of the magnetic $Fe_3O_4$@$SiO_2$ nanorods, which is consistent with the DOP changes in **Figure 2d**. In the case of hybrid nanorods, the strong resonance at λL changes the polarization from being horizontal to vertical under the scenario of horizontally aligned hybrid nanorods (**Figure 2e** and **Figure S3e**), leading to an orthogonal polarization effect.

The studies on the polarization states of the transmitted light indicate that assembly of hybrid nanorods into nematic liquid crystal phases leads to wavelength-dependent polarization effects. At λT, unpolarized incident light became polarized parallel to the direction of the nanorod alignment, as shown schematically in **Figure 2g**. At λL, the polarization direction of the transmitted light is perpendicular to the nanorod alignment, as shown schematically in **Figure 2h**. This correlation between polarization and the plasmonic absorption peaks can be explained by anisotropic electron–photon interactions in nanorods. Transverse plasmon resonance is induced by the resonant oscillation of free electrons along the rods' short axes, which absorb light that has a component of its electric field vector perpendicular to the rod alignment. As a result, the selective absorption of perpendicularly polarized light converts unpolarized light into light that is polarized parallel to the nanorod alignment. In contrast, the longitudinal plasmon resonance is induced by the resonant oscillation of free electrons along the long axes of the nanorods, leading to the transmitted light being polarized perpendicular to the nanorod alignment. Such directional photon–electron interactions explain the wavelength-dependent polarization effects and can be used to design interesting patterns by stacking plasmonic films at different angles. For example, the images in **Figure 2i** demonstrate that the absorption-induced polarization of transmitted light of one plasmonic film can induce selective plasmonic excitation of a second film, with the excitation modes depending on twisting angles. As a result, we observe the presence of patterns when stacking a plain plasmonic film on a patterned plasmonic film, as shown in the bottom panels in **Figure 2i**. Changing the twist angle from $0^{o}$ to $90^{o}$ switches the color of the central domain from red to blue, suggesting a transition from selective transverse-mode excitation (red color) to longitudinal-mode excitation (blue color).

**Twist-angle-dependent plasmonic excitation in moiré plasmonic superlattices**

To explain the correlated plasmonic excitation, we measured concentration-dependent extinction spectra of the hybrid nanorods in different phases, including random phases without a magnetic field and nematic liquid-crystal phases under parallel, perpendicular, and vertical magnetic fields. As shown in **Figure 3a**, in the absence of a magnetic field, the absorption increases proportionally to the concentration of the hybrid nanorods. In the presence of a vertically oriented magnetic field, the nematic liquid-crystal phase exhibits a different dependence on concentration; there is a consistent increase in absorption at λT and a gradual disappearance of the absorption peaks at λL (**Figure 3b** and **Figure S4a**). Application of a magnetic field along the direction of propagation of the light (**Figure S4b**) enhances absorption by the transverse mode and suppresses absorption by the longitudinal mode. The detailed dependences of the absorption at λT and λL are shown in **Figures 3c** and **d**, respectively. In all different samples, we observed a linear increase of absorption at λT with the concentration of the hybrid nanorods (**Figure 3c**). However, as shown in **Figure 3d**, we observed a nonlinear dependence of absorption at λL on hybrid nanorod concentration when the nanorods were aligned perpendicularly to the incident direction of light under either a vertical or horizontal magnetic field. A close analysis indicates that the slope of the dependences decreases at high concentrations, which suggests a possible complete extinction of incident light of a particular polarization state at the longitudinal mode. That is, above the concentration threshold, all the light components that have electric field vectors that are parallel to the nanorods' alignment are absorbed completely due to longitudinal plasmon resonance. Such complete extinction is responsible for the nonlinear behavior of the extinction intensity as concentration increases. To verify this scenario, we investigated the polarization states of the transmitted light of these solutions using the rotating quarter-wave plate Stokes polarimetry. As demonstrated in **Figure 3e**, for aligned nanorods, the DOP of light at λT increases linearly with concentration at concentrations below 0.33 mg/mL, whereas the DOP at λL approaches a plateau and then saturates. The polarization direction of the measurement is consistent with our previous analysis on the directional free-electron resonance in the nanorods. This sequence of events demonstrates that, at λL, light with its electric field oscillating along the long axes of the nanorods is absorbed completely at high concentrations, leading to the saturation of the DOP. The extinction of the multicomponent $Fe_3O_4$/Au@RF nanorods has three main contributions: the plasmonic resonance of Au nanorods, the absorption of $Fe_3O_4$, and the absorption of RF. Only the absorption induced by plasmonic resonance of the Au nanorods depends on the polarization of the light. As a result, the contribution of the Au nanorods to the absorption at λL reaches a maximum and does not increase at concentrations above the threshold (0.33 mg/mL), whereas the absorptions of $Fe_3O_4$ and RF keep growing because they are not polarization-dependent. Therefore, the absorption peak of Au nanorods at λL eventually merges into the broadband absorption of the $Fe_3O_4$ and RF, which explains the disappearance of the longitudinal peak of the Au nanorods at high concentrations in **Figure 3b**. Based on this analysis, we conclude that absorption of nematic plasmonic superlattices is directional, and the longitudinal mode reaches saturation status above the threshold concentration of 0.33 mg/mL. The distinct absorption profiles of the transverse and longitudinal modes exhibit subtle color changes of aligned nanorods under non-polarized irradiation. As shown in **Figure 3f**, films show consistent red and green colors under irradiation by polarized light, due to selective excitation of the transverse and longitudinal modes, respectively. The perceived color changes from light green at low concentration to dark red at high concentrations under unpolarized irradiation. At low concentrations, the absorption is dominated by the longitudinal mode (see peak position in **Figure S1**), leading to the complementary green color. Once the threshold concentration of 0.33 mg/mL is reached, the absorption begins to be dominated by the transverse mode, leading to the complementary red color. As shown in **Figure 3g**, at λT, the polarization state changes from unpolarized to vertically polarized with ψ of 90° (**Figure S5**) and a DOP of 32.3% in thick films (**Figure 3h**). The transition is shown in the inset Poincaré sphere in **Figure 3h**, demonstrating the gradual rotation of the polarization direction to the final vertically polarized state as the film thickness increases.

**Mapping polarization states and moiré plasmonic phases**

Based on the above polarization experiments, we hypothesize the existence of an extinction-induced plasmonic excitation correlation in the twisted nematic plasmonic superstructures. At a twist angle of 0°, the nanorod alignment is parallel in the two films. Hybrid nanorods with unidirectional alignment absorb light at λL to such an extent that light with parallel polarization is completely absorbed at small twist angles, leading to the saturation of DOP at 100% in **Figure 3e**. As a result of longitudinal-mode saturation, the absorption spectrum is dominated by the transverse mode, with a strong peak at λT at 0° (**Figure 4a**). Changing the twist angle to 90° makes the nanorods perpendicularly aligned in the two stacked films, leading to maximum absorption of light with its polarization along either of two orthogonal directions. Absorption by the longitudinal mode cannot readily saturate at large twist angles, leading to a dominant longitudinal peak (**Figure 4a**). To elucidate the twist-induced plasmonic excitation correlation and complete extinction at λL, we measured the DOP at different twist angles. **Figure 4b** shows the DOP at λL as a function of the twist angle. There is a gradual increase of the DOP up to 100% as the twist angle goes from 90° to 0°. The corresponding transition of the polarization ellipse in **Figure 4c** demonstrates a nearly linear polarization at 0o (perpendicular to nanorod alignment), which is consistent with the relationship between nanorod alignment and the polarization of the transmitted light.

To understand the moiré patterns in **Figures 1g** and **h**, we used rotating quarter-wave-plate Stokes polarimetry to map the light polarization states induced by transmission through the twisted nematic plasmonic superlattices. Based on the same measurement scheme in **Figure 2a**, the twisted plasmonic films shown in **Figures 1g** and **h** were placed before the quarter-wave plate, and a digital camera was used as the detector. For a moiré lattice at each given twist angle, we took nine images of the twisted films by changing the fast axis of the quarter-wave plate from $0^{o}$ to $180^{o}$ in $20^{o}$ increments and observed the intensity changes in the moiré patterns (**Figures S6** and **S7**). An algorithm was developed to derive the intensity corresponding to each pixel in the images. Analysis of the transmitted intensity change as a function of the quarter-wave plate angle enables us to calculate the light polarization state at each pixel with high spatial resolution and throughput. Through the pixel-wise data processing, we can map polarization states using the DOP, the orientation angle ($\psi$), and the ellipticity angle ($\chi$) of the transmitted light as a function of position. For illumination at $\lambda L$, we observed well-defined DOP domains that underwent consistent changes as the twist angle increased to $90^{o}$ (**Figure 4d** and **Figure S8a**). Although DOP mapping outlines domains with the same degree of polarization, $\psi$ mapping can resolve the orientation difference even in domains with the same DOP, leading to a different and more defined characterization of the polarization state, as demonstrated in **Figures 4e** and **Figure S8b**. At a twist angle of $45^{o}$, we observed more complicated patterns in $\psi$ mapping compared to those seen in DOP mapping, because $\psi$ mapping can resolve finer structures having the same DOP but different polarization orientations. The $\chi$ mapping indicates the spatial dependence of the ellipticity angle of the transmitted light (**Figure S8c**) and demonstrates nearly linear polarization. For illumination at $\lambda T$, the change of the DOP with the twist angle is less obvious than that for illumination at $\lambda L$ (**Figures 4f** and **Figure S9a**). However, we still managed to identify domains with different DOPs. In contrast, the $\psi$ (**Figures 4g** and **Figure S9b**) and $\chi$ (**Figure S9c**) mappings at various twist angles are not well-defined, most likely due to small changes in polarization states at the transverse mode.

**Reconfigurable plasmonic moiré superlattices**

The interaction between light polarization and orientation-dependent plasmonic excitation coupled in the twisted nematic plasmonic bilayer enables the creation of reconfigurable moiré superlattices. One benefit associated with the hybrid nanorods is magnetic orientation control, which enables the preparation of patterned plasmonic films with nanorods being aligned distinctly in different defined domains. We prepared two plasmonic films with identical, one-dimensional, periodic stripes (**Figure 5a** and **Figure S10**) in which the hybrid nanorods were alternately aligned in the plane of the film along perpendicular directions in neighboring domains (**Figures S10a** and **b**), leading to the expected green and red colors upon illumination with light that was polarized parallel to the stripes. The stripes are not visible upon illumination with unpolarized light (**Figure S10a**). Upon the translation operation shown in **Figure 5b**, in which the two plasmonic films were first overlapped at (0, 0), and then the top layer was slid along the y-axis (i.e., perpendicular to the stripe direction) to the coordinate (0, 1), where the units are the width of a single stripe, we observed a gradual transition from red to blue in the stacked areas (**Figure 5c** and **Supporting Video 1**). At (0, 0), the nanorod alignment in each stripe in the adjacent films in the bilayer is parallel. Because the transverse extinction is stronger and is less sensitive to polarization than the longitudinal extinction, the bilayer takes on a nearly regular red color across the entire stacked area. As the top layer is slid to (0, 1/6), narrow, periodic blue domains appear between adjacent red domains. These blue domains correspond to regions in which the nanorods have orthogonal orientations in the bilayer, enhancing the longitudinal extinction. Further increasing the displacement to one stripe width causes the entire bilayer to be blue, due to the saturation of extinction in the transverse peak. This color transition is consistent with the relative alignment changes in the stacked plasmonic bilayer (**Figure S10c**), leading to complete color switching from red to blue in the moiré superlattice.

In the rotational operation (**Figure 5d**), the moiré superlattice in the twisted bilayer is highly dependent on the twist angle, leading to complex patterns and color changes. We investigated twist angles ranging from $0^{o}$ to $90^{o}$. As the twist angle increases to $90^{o}$, the perceived patterns change from 1D stripes to 1D triangular lattices, 2D rhomboidal lattices, and finally 2D square lattices (**Figure 5e** and **Supporting video 2**), all featuring alternating blue and red domains. This transition is induced by the alignment changes of the nanorods in the stacked regions of the two layers, with parallel and perpendicular alignment leading to red and blue colors, respectively. Changing the symmetry and dimension of the constituent plasmonic films creates even more complicated moiré patterns, with unique features and symmetries that are not attainable from conventional lithographic or assembly methods. For instance, we prepared two plasmonic films with 2D circle lattices (**Figure S11**) and observed the resulting moiré pattern formation. Moving the top layer along the x-axis yields a continuous transition of the moiré superlattices that is highly correlated with the nanorod alignment and is also reconfigurable depending on stacking angle (**Figure 5f**). Notably, moiré superlattices with perpendicular stacking share the same patterns and symmetries as the superlattices with parallel stacking, albeit with opposite colors in the corresponding domains (**Figures S12a** and **b**). Following the rotational operation, the moiré superlattices demonstrate non-classical patterns with both translational and rotational symmetries. As shown in **Figure 5g**, starting from a parallel stacking ($0^{o}$), the moiré superlattices have the same patterns and contrast for both clockwise (bottom panels) and counterclockwise twisting (top panels), which is in good agreement with the twist-direction-independent plasmonic excitation in the twisted bilayer. These bilayers are periodic superstructures comprising unit cells of different symmetries and length scales. Notably, the moiré patterns can only be observed in the stacked areas. The non-stacked regions remain highly transparent and devoid of noticeable features.

Stacking two films containing nematic domains of different symmetries and dimensions can generate additional, unique moiré patterns. As demonstrated in **Figure S13**, when one plasmonic film with a 1D stripe is stacked over a second plasmonic film with 2D circle lattices, 2D moiré superlattices with both linear and circular features are produced. The inherent mismatch in periodicity, symmetry, and dimension between the twisted bilayer leads to gradient changes in domain colors, shapes, and sizes.

**Conclusions**

This work reports highly reconfigurable and complex moiré superlattices in twisted plasmonic bilayers with pre-defined nematic domains of hybrid $Fe_3O_4$/Au nanorods and recognizes the correlated light–matter interactions between the collective polarization effect in nematic plasmonic superstructures and the orientation-dependent plasmonic extinction of the individual nanorods. The systematic studies with carefully designed experiments demonstrate remarkable twist-angle-dependent plasmonic extinction enhancement in the twisted bilayer, which is accompanied by color changes in response to both translational and rotational operations. The nematic phase of hybrid nanorods causes unpolarized light that impinges on these materials to become polarized, with the polarization direction being perpendicular and parallel to the nanorods' alignment for the longitudinal and transverse modes, respectively. Due to the strong longitudinal plasmon resonance, the longitudinal-mode absorption reaches a saturated state above a threshold concentration, and therefore the transverse mode absorption dominates the absorption spectrum, leading to a thickness-dependent, correlated, transverse-mode enhancement. This absorption saturation explains the correlated plasmonic excitation and the observed continuous transition of polarization states as the twist angle increases from $0^{o}$ to $90^{o}$. This fundamental understanding enables the design of complicated superlattices with highly reconfigurable and defined symmetries, domain colors, feature sizes, and contrast gradients. Although the feature symmetry is highly correlated with the initial pattern geometry encoded in each plasmonic sub-lattice, the two-dimensional twisted superlattices have theoretically unlimited patterns and appearances corresponding to each possible twisting and stacking configuration. Light–matter interactions in these nematic plasmonic phases and the twisted bilayers open the door to developing topological, chiral optics and twist-optics, providing the opportunity to create advanced optical and optoelectronic materials.

ASSOCIATED CONTENT

**Supporting Information**. Supporting information files, including absorption spectra, polarization ellipse, digital pictures of photonic Moiré lattices, and Moiré phase mapping. This material is available free of charge via the Internet at http://pubs.acs.org.

## AUTHOR INFORMATION

### Corresponding Author

Zhiwei Li - Department of Chemistry and Biochemistry, University of Maryland, College Park, Maryland 20742, USA; Email: zlinano1@umd.edu

### Authors

Rui Huang - Department of Chemistry and Biochemistry, University of Maryland, College Park, Maryland 20742, USA.

Jordan Austin-Frank Wilson - Department of Chemistry and Biochemistry, University of Maryland, College Park, 20742, USA.

Anders Dollard - Department of Chemistry and Biochemistry, University of Maryland, College Park, 20742, USA.

Nicholas C Fisher - Department of Chemistry and Biochemistry, University of Maryland, College Park, 20742, USA.

Zixian Fang - Department of Chemistry and Biochemistry, University of Maryland, College Park, 20742, USA.

John T. Fourkas - Department of Chemistry and Biochemistry, University of Maryland, College Park, 20742, USA.

### Author Contributions

The manuscript was written through the contributions of all authors. All authors have given approval to the final version of the manuscript. ‡These authors contributed equally.

### Funding Sources

The authors gratefully acknowledge financial support from the University of Maryland, College Park.

### Notes

The authors declare no competing financial interests.

ACKNOWLEDGMENT

We acknowledge the support of the Maryland NanoCenter and its AIMLab. This work was supported by the startup funds from the University of Maryland, College Park.

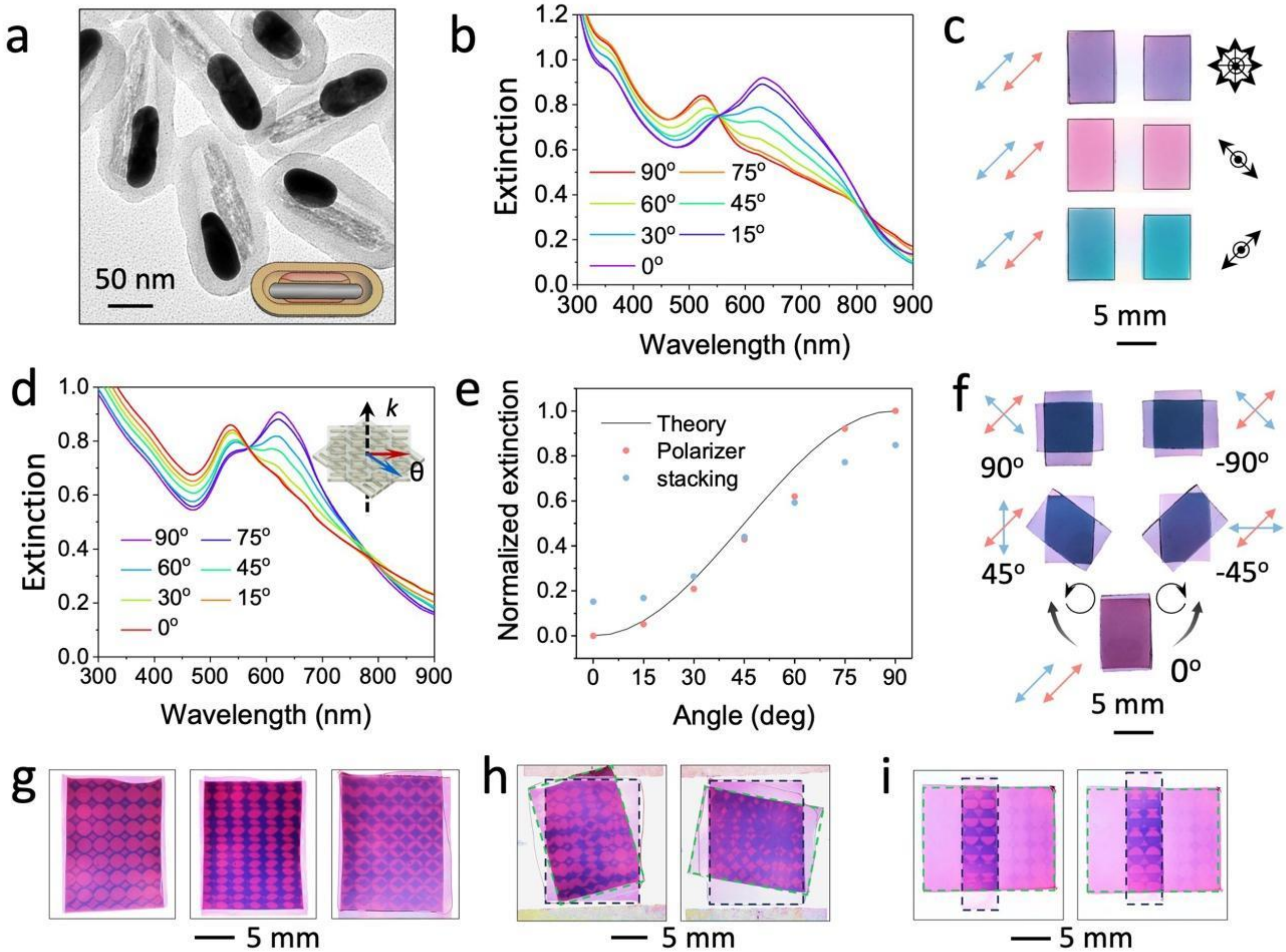

**Figure 1. Correlated plasmonic excitation in twisted plasmonic superstructures.** (**a**) TEM image of the hybrid $Fe_3O_4$/Au@RF nanorods. The inset shows a schematic of the hybrid nanorods. (**b**) Extinction spectra for a plasmonic film composed of parallel hybrid nanorods irradiated with linearly polarized light as a function of the angle between the polarization and the nanorod alignment axis. (**c**) Optical images of two identical plasmonic films irradiated with unpolarized (top), -45° polarized, and 45° polarized light. The rod alignments in the left and right panels are illustrated with blue and red arrows, respectively. The polarization of incident light is indicated by black arrows. (**d**) Extinction spectra of a clockwise-twisted plasmonic bilayer under excitation of unpolarized light at different twist angles (α), i.e., the angles between the alignment of two different directions in the inset. (**e**) Dependence of the normalized extinction at longitudinal mode (λL) on rod alignment relative to the polarization for a single film irradiated with linearly polarized light (red dots) and on the stacking angle of a twisted plasmonic bilayer irradiated with unpolarized light (blue dots). (**f**) Optical images of plasmonic bilayers at different twist angles. The rod alignments in the stacked top and bottom layers are indicated by blue and red arrows, respectively. Optical images of Moiré patterns in stacked plasmonic superstructures twisted by (**g**) 0° and (**h**) 15° (left) and 75° (right). (**i**) Optical images of AB stacking of plasmonic superstructures of different patterns.

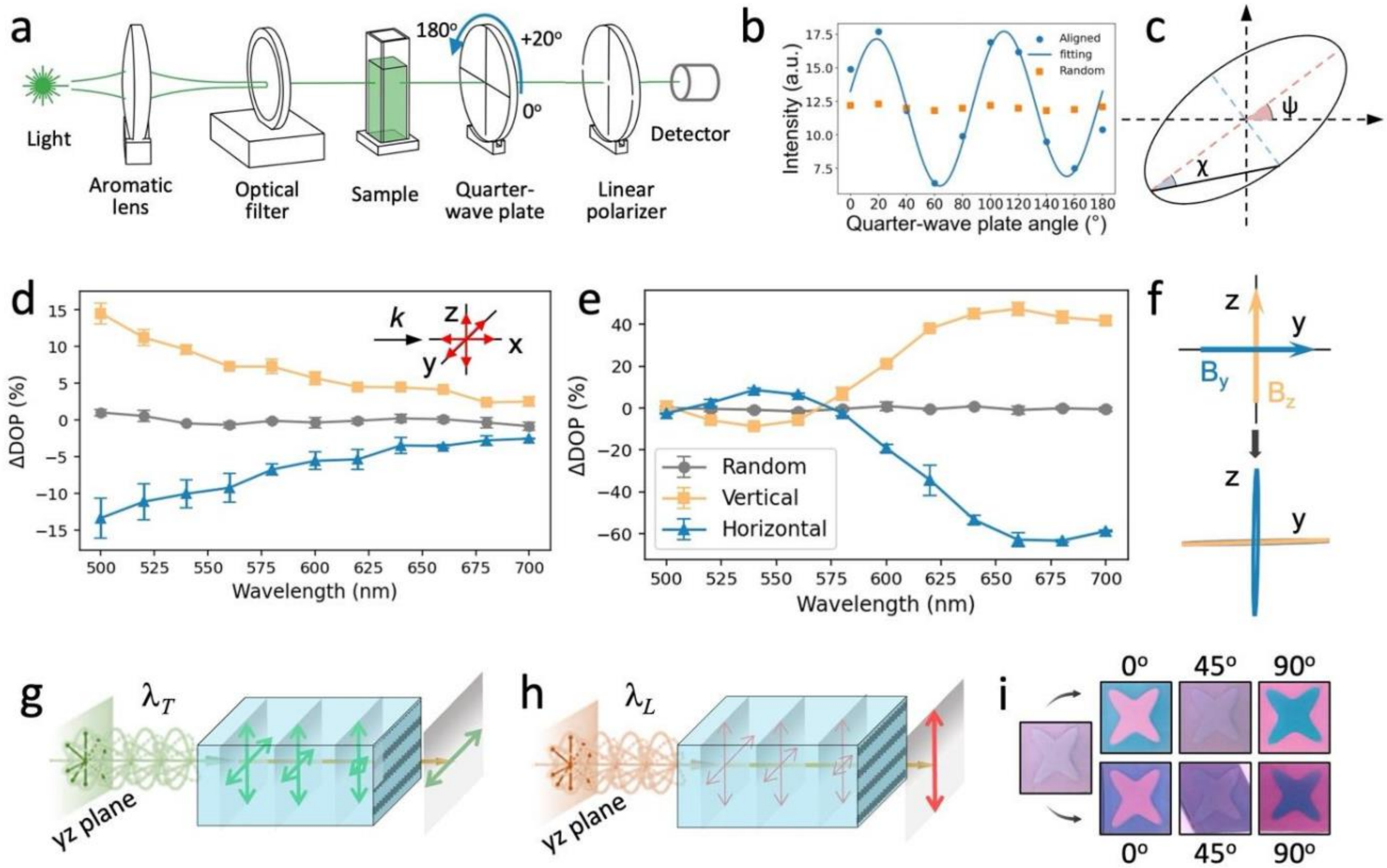

**Figure 2. Localized-surface-plasmon-resonance-induced polarization in liquid-crystal plasmonic superstructures.** (**a**) Schematic illustration of the rotating quarter-wave plate Stokes polarimetry to measure the light polarization states. (**b**) Dependence of light intensity on the angles of the quarter-wave plate during the measurements. (**c**) A schematic illustration of polarization ellipse representation, in which orientation angle ψ and ellipticity angle χ are introduced to define light polarization states. Changes in the DOP induced by liquid crystals made of (**d**) magnetic nanorods and (**e**) hybrid nanorods. The inset shows parallel, horizontal, and vertical magnetic fields being applied along the x-, y-, and z-axes, respectively. (**f**) Schematic illustration of vertical (yellow arrow) and horizontal (blue arrow) magnetic fields being applied along the z- and y-axes, respectively. The resulting polarization states under the vertical and horizontal fields are represented by the polarization ellipses of the same color in the bottom panel. Schematic illustration of changes in light polarization in the plasmonic superstructures at (**g**) λT and (**h**) λL. (**i**) Optical images of plasmonic liquid crystals with a polarizer (top) and with a second plasmonic liquid-crystal film (bottom) at the designated twisting angles.

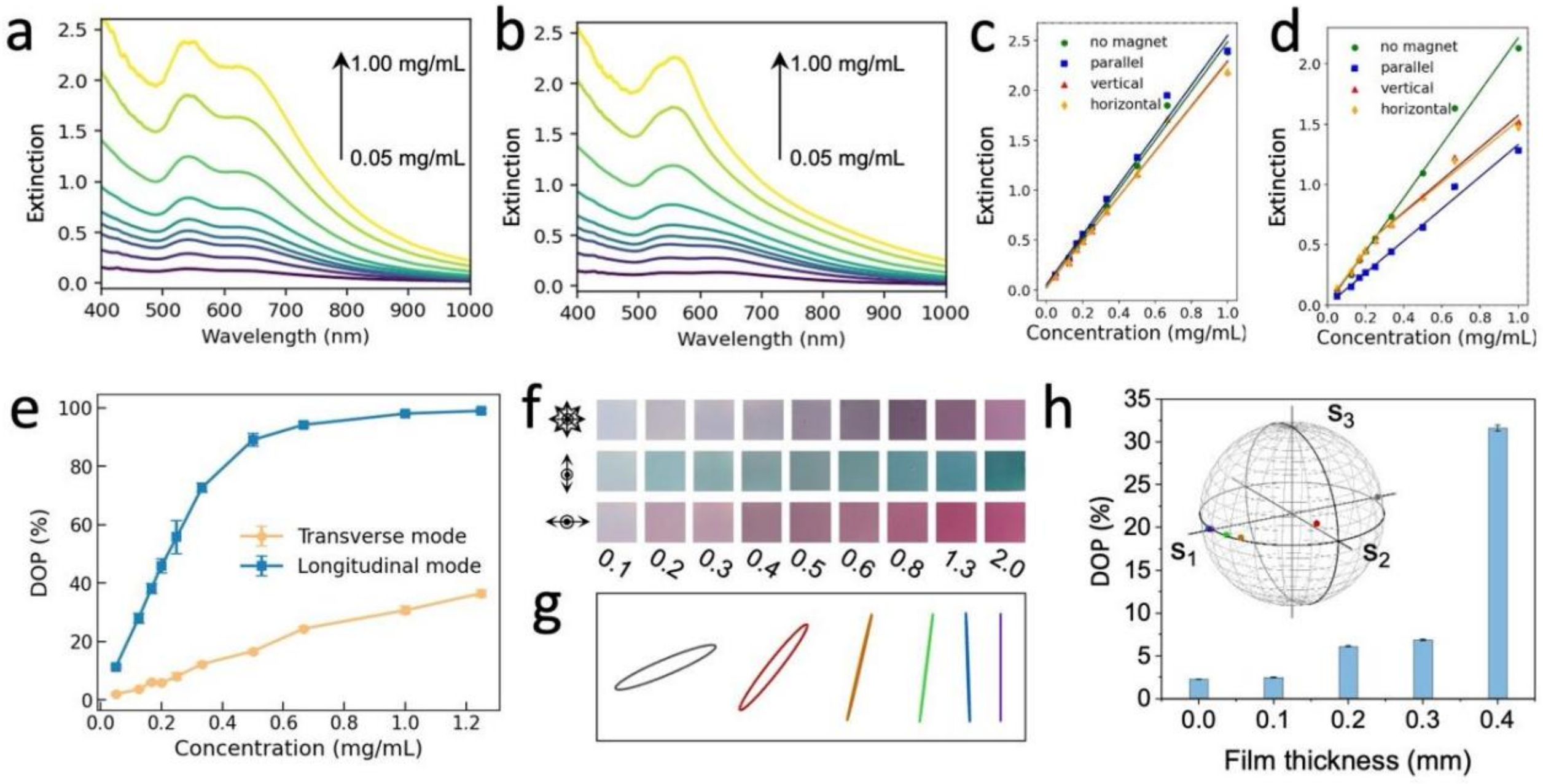

**Figure 3.** Extinction spectra of a hybrid nanorod solution with (**a**) no magnetic field and (**b**) a vertical magnetic field. The concentration of the samples increases from 0.05 mg/mL to 0.13 mg/mL, 0.17 mg/mL, 0.20 mg/mL, 0.25 mg/mL, 0.33mg/mL, 0.50 mg/mL, 0.67 mg/mL, and 1.00 mg/mL. The dependence of the peak extinction of the hybrid nanorods at (**c**) lT and (**d**) lL under no, parallel, vertical, and horizontal magnetic fields. (**e**) The degree of polarization of hybrid nanorod dispersions of different concentrations under unpolarized light excitation. The nanorods are magnetically oriented perpendicular to the incident direction of the light. (**f**) Optical images of liquid-crystal films of different thicknesses made by vertically aligning the hybrid nanorods. (**g**) Polarization-ellipse representation of the polarization state of transmitted light through the liquid-crystal films of different thicknesses. (**h**) DOP for different film thicknesses. Inset: the corresponding Poincaré spheres representing the polarization state changes.

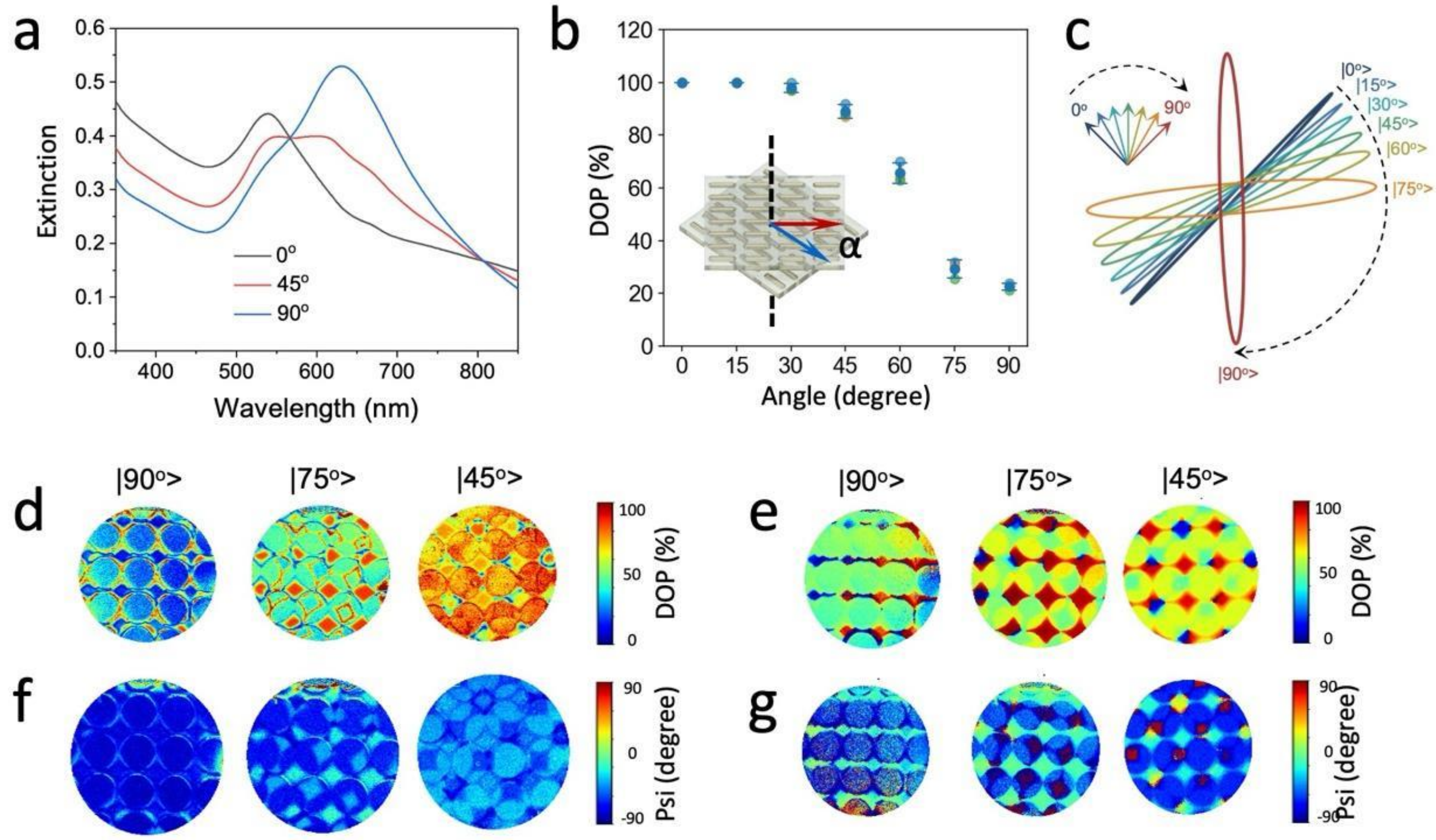

**Figure 4. Mapping the correlated moiré phases in twisted plasmonic superstructures.** (**a**) Extinction spectra of a plasmonic film when another identical film was used to calibrate the spectrometer. (**b**) Dependence of DOP on twisting angles of the twisted plasmonic liquid crystals under excitation of unpolarized light. Inset: scheme of the twisted plasmonic liquid crystals. (**c**) Twist-angle-dependent polarization states of the twisted plasmonic bilayers represented by the polarization ellipses. The inset shows the changes of the nanorods' twist angle in the top layer from 0° to 90° while the nanorods' alignment is fixed at 0° in the bottom layer. Moiré phase mapping of the twisted plasmonic films in terms of (**d**) the degree of polarization and (**e**) the polarization angle at $\lambda$L. Moiré phase mapping of the twisted plasmonic films in terms of (**f**) the degree of polarization and (**g**) the polarization angle at $\lambda$T.

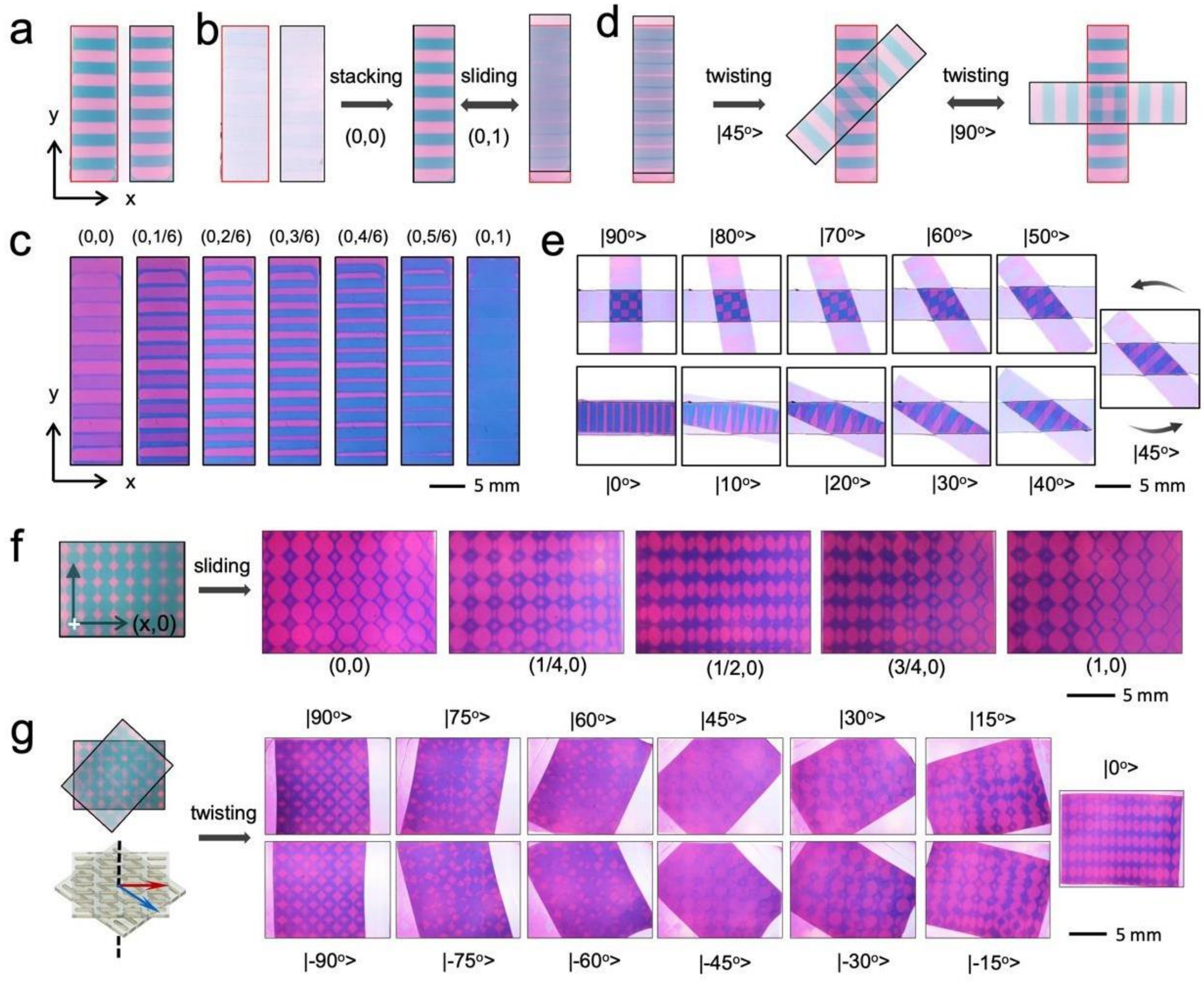

**Figure 5. Plasmonic moiré patterns in reconfigurable twisted plasmonic superstructures.** (**a**) Optical images of two plasmonic films with well-defined 1D stripes under polarized illumination. (**b**) Schematic illustration of the stacked bilayer under a translational operation and (**c**) optical images of the stacked bilayer at different displacements. One displacement unit is the width of a stripe. (**d**) Schematic illustration of the twisted bilayer under a rotational operation and (**e**) the corresponding optical images of the twisted bilayer at different twist angles. (**f**) Schematic illustration (left panel) of two plasmonic films with pre-designed 2D circular patterns and optical images (right panels) showing the Moiré superlattices under different displacements (see labels). (**g**) Schematic illustration (left panel) of two plasmonic films and optical images (right panels) showing the reconfigurable Moiré superlattices with different twist angles (see labels).

**Supplementary Information**

# Correlated Plasmonic Excitation in Twisted Nematic Plasmonic Superlattices

Rui Huang[‡,1], Jordan Austin-Frank Wilson[‡,1], Anders Dollard[1], Nicholas C Fisher[1], Zixian Fang[1], John T. Fourkas[1,2,3], Zhiwei Li*[,1]

[1]Department of Chemistry and Biochemistry, University of Maryland, College Park, Maryland 20714, United States

[2]Institute for Physical Science and Technology, University of Maryland, College Park, Maryland 20714, United States

[3]Maryland Quantum Materials Center, University of Maryland, College Park, Maryland 20714, United States

Email: zlinano1@umd.edu

‡ These authors contributed equally.

## Characterization

Transmission electron micrographs of the hybrid nanorods were obtained with a JEM 2100 LaB6 transmission electron microscope operating at 120 kV. Optical images were obtained using a cellphone camera. Ultraviolet–visible–near-infrared (UV-Vis-NIR) spectra were measured using an Ocean Optics HR4000 CG-UV-NIR high-resolution spectrometer.

## Calculation of the plasmonic extinction in the twisted plasmonic bilayer

The plasmonic extinction of the transverse and longitudinal modes of Au nanorods can be described by the following expressions:

$$\frac{\langle A_L(\omega)\rangle - \langle A_L(90°)\rangle}{\langle A_L(0°)\rangle - \langle A_L(90°)\rangle} = \cos^2\omega \qquad \text{(S1)}$$

and

$$\frac{\langle A_T(\omega)\rangle - \langle A_T(0°)\rangle}{\langle A_T(90°)\rangle - \langle A_T(0°)\rangle} = \sin^2\omega\ , \qquad \text{(S2)}$$

where ω is the angle between the rod alignment and the light polarization in plane of the nanorods and $< A_L(\omega) >$ is the expectation value of the absorbance of the longitudinal mode for light polarized at an angle of ω. The subscripts $L$ and $T$ represent the longitudinal and transverse modes, respectively. Using these equations, the extinction strength of longitudinal and transverse modes can be expressed as $\cos^2 \omega$ and $\sin^2 \omega$, respectively. When $\omega = 0^o$ (i.e., when the rod alignment is parallel to the light polarization), the longitudinal mode is excited selectively, with a maximum normalized extinction strength of 1, whereas extinction by the transverse mode is completely suppressed, giving a normalized extinction strength of 0. This theoretical consideration is consistent with the spectra of hybrid nanorods measured under linearly polarized light and a changing magnetic field (**Figure 1b**). Furthermore, when the extinction at $\lambda_L$ measured under different magnetic fields is substituted into Eq. S1, and the extinction strength can be extracted from experimental measurements. More specifically, the absorbance of the longitudinal mode, $< A_L(\omega) >$, is measured from the spectra at different orientation angles (ω). Substituting the absorbance at angles of ω, 0°, and 90° into Eq. 1 produces the relative extinction strength. The calculation of extinction strength of transverse mode was based on Eq. 2. As shown by the red dots in **Figure 1e**, such values are in good agreement with the theoretical values for different polarizations.

Using the same method, it is possible to derive the extinction strength in the twisted plasmonic bilayer at different twist angles. Specifically, the extinction of the twisted plasmonic bilayer can be expressed as the superposition of the extinction of Au nanorods in the two films, and the constant extinction from $Fe_3O_4$ and polymer shells in the two films. Therefore, the longitudinal and transverse extinction in the twisted plasmonic bilayer at a twist angle of α can be calculated from

$$< A(\alpha) > \; = \; < A(\alpha) >_1 + < A(\alpha) >_2 \tag{S3}$$

where

$$< A(\alpha) >_1 = < \psi_L|\mathrm{P}|\psi_L >_1 + C_1 \tag{S4}$$

and

$$< A(\alpha) >_2 = < \psi_L|\mathrm{P}|\psi_L >_2 + C_2\,. \tag{S5}$$

Here, the subscripts 1 and 2 indicate the first film and second film, respectively. The term $<\psi_L|\mathrm{P}|\psi_L>$ represents the expectation absorbance of Au nanorods at longitudinal mode. The ψ means polarization function of the plasmonic liquid crystals and P is the polarization operator. Under unpolarized light incidence, the extinction from the first film ($< A(\alpha) >_1$) and the extinction from the $Fe_3O_4$ nanorods and polymer shells in the second film ($C_2$) should be constant for different stacking angles. Only the extinction

of Au nanorods from the second film ($<\psi_L|\mathrm{P}|\psi_L>_2$) changes accordingly in different stacking angles due to the polarization effect of the first film. Eqs. S3 to S5 are valid for both the transverse and longitudinal modes. We used the longitudinal mode for deriving the extinction strength in the twisted plasmonic films at different stacking angles. Because the longitudinal plasmonic mode is not saturated, it is reasonable to assume that the longitudinal plasmonic extinction is equivalent to the plasmonic excitation at an angle of ω under linearly polarized light. Therefore, the longitudinal plasmonic excitation at a stacking angle of 90$^o$ and 0$^o$ can be calculated as

$$\begin{aligned} &< \mathrm{A}(90°) > - < \mathrm{A}(0°) > \\ &=< \mathrm{A}(90°) >_2 - < \mathrm{A}(0°) >_2 \\ &= \mathrm{I_L}\sin^2\omega - \mathrm{I_L}\sin^2(90° - \omega)\,, \end{aligned} \tag{S6}$$

where $I_L$ is the maximum longitudinal plasmonic excitation. Combining Eqs. S1 and S6 leads to:

$$\frac{<\mathrm{A}(90°)>-<\mathrm{A}(0°)>}{<\mathrm{A_L}(0°)>-<\mathrm{A_L}(90°)>} = \sin^2\omega - \sin^2(90° - \omega)\,, \tag{S7}$$

where the numerator is the difference between the peak extinctions at twist angles of 90$^o$ and 0$^o$ and denominator is the difference between the peak extinction of the plasmonic film when the incorporated Au nanorods are parallel and perpendicular to the light polarization direction. Solving this equation yields the equivalent extinction angle (ω) at a twist angle of 90$^o$. The extinction strengths at stacking angles of 90$^o$ and 0$^o$ are $\sin^2\omega$ and $\sin^2(90° - \omega)$, respectively. The extinction strength for different twist angles can be calculated using the same method (blue dots in **Figure 1e**).

**Measurement of light polarization states using rotating quarter-wave-plate Stokes polarimetry**

Rotating quarter-wave plate Stokes polarimetry is used in this work to measure the polarization states of light. The scheme of the measurement is shown in **Figure 2a**. The light intensity after transmission through a sample is by rotating the quarter-wave plate from 0$^o$ to 180$^o$ in 20$^o$ increments. At 0$^o$, the fast axis of the quarter-wave plate is parallel to the polarization direction of the polarizer. The twist angle (θ)-light intensity scatter plot is fitted using

$$I(\theta) = \frac{1}{2}\left[S_0 + \frac{S_1}{2}(1 + cos4\theta) + \frac{S_2}{2}sin4\theta - S_3 sin2\theta\right] \tag{S8}$$

to derive the four Stokes parameters ($S_0$, $S_1$, $S_2$, $S_3$) of the light. Based on the fitting coefficients, the Stokes parameters can be calculated and plotted in the Poincaré spheres representation to show the light polarization states after normalizing $S_1$, $S_2$, and $S_3$ against $S_0$. The orientation angle (azimuth, ψ) and ellipticity angle ( χ ) can be calculated using:

$$\psi = \frac{1}{2} tan^{-1}\left(\frac{S_2}{S_1}\right) \tag{S9}$$

and

$$\chi = \frac{1}{2} sin^{-1}\left(\frac{S_3}{S_0}\right) . \tag{S10}$$

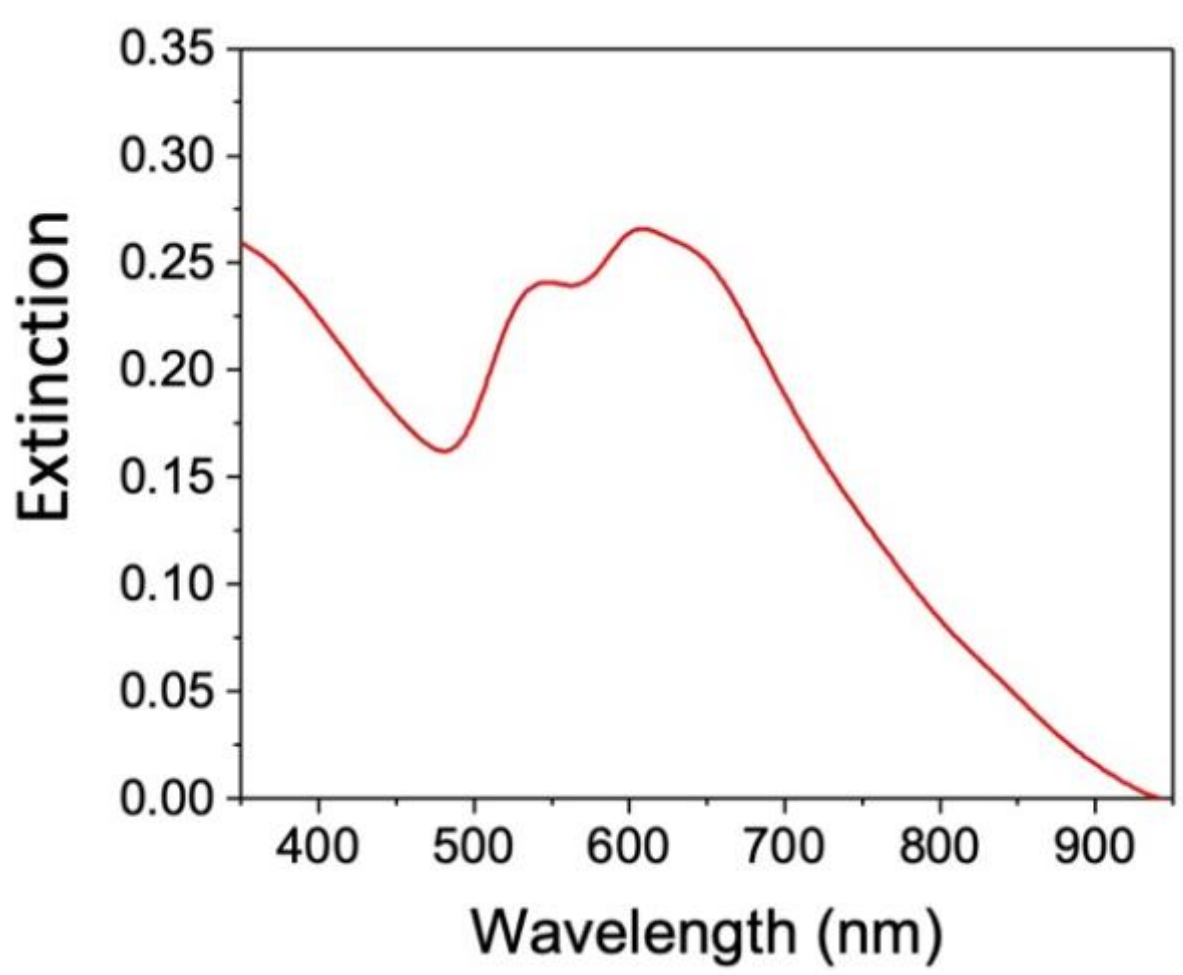

**Figure S1.** Extinction spectra of the hybrid nanorods.

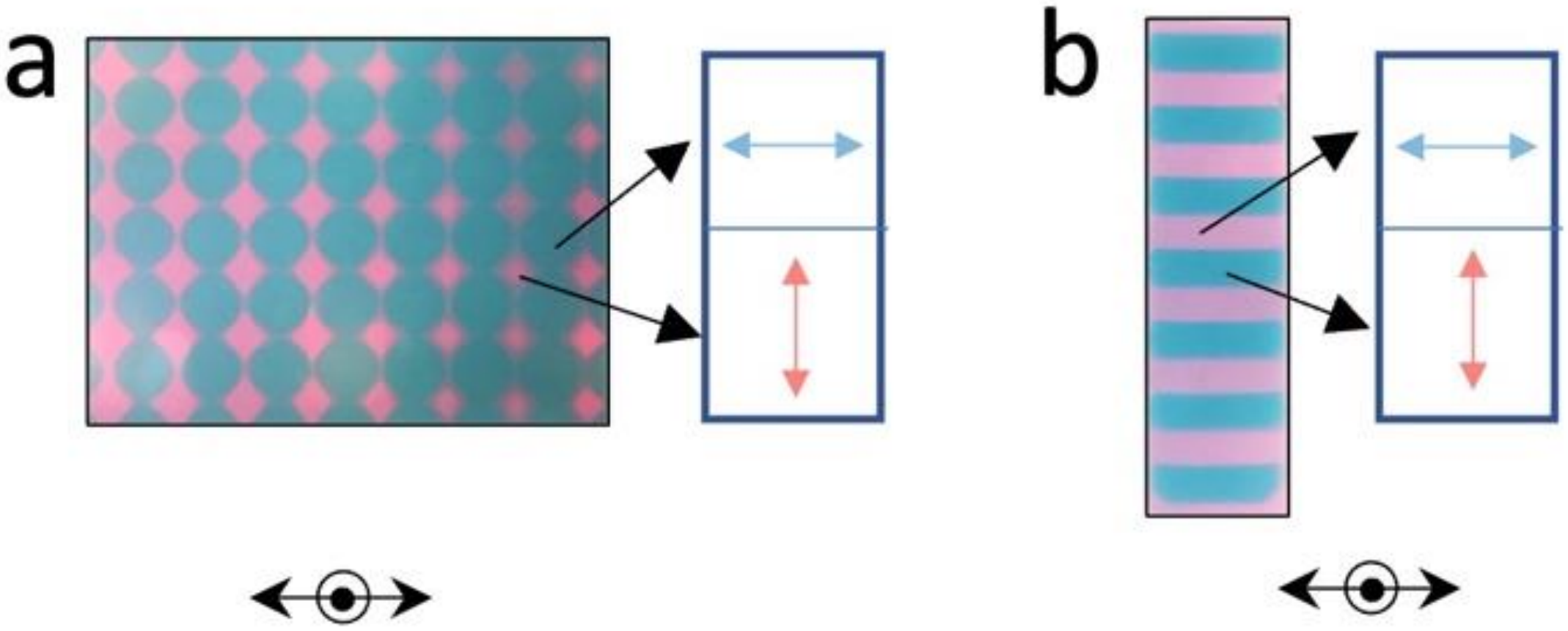

**Figure S2.** Optical images of plasmonic patterns of **(a)** circles and **(b)** stripes created by magnetically aligning the hybrid nanorods vertically and horizontally in different domains. The red and blue arrows indicate vertically and horizontally aligned nanorods, respectively.

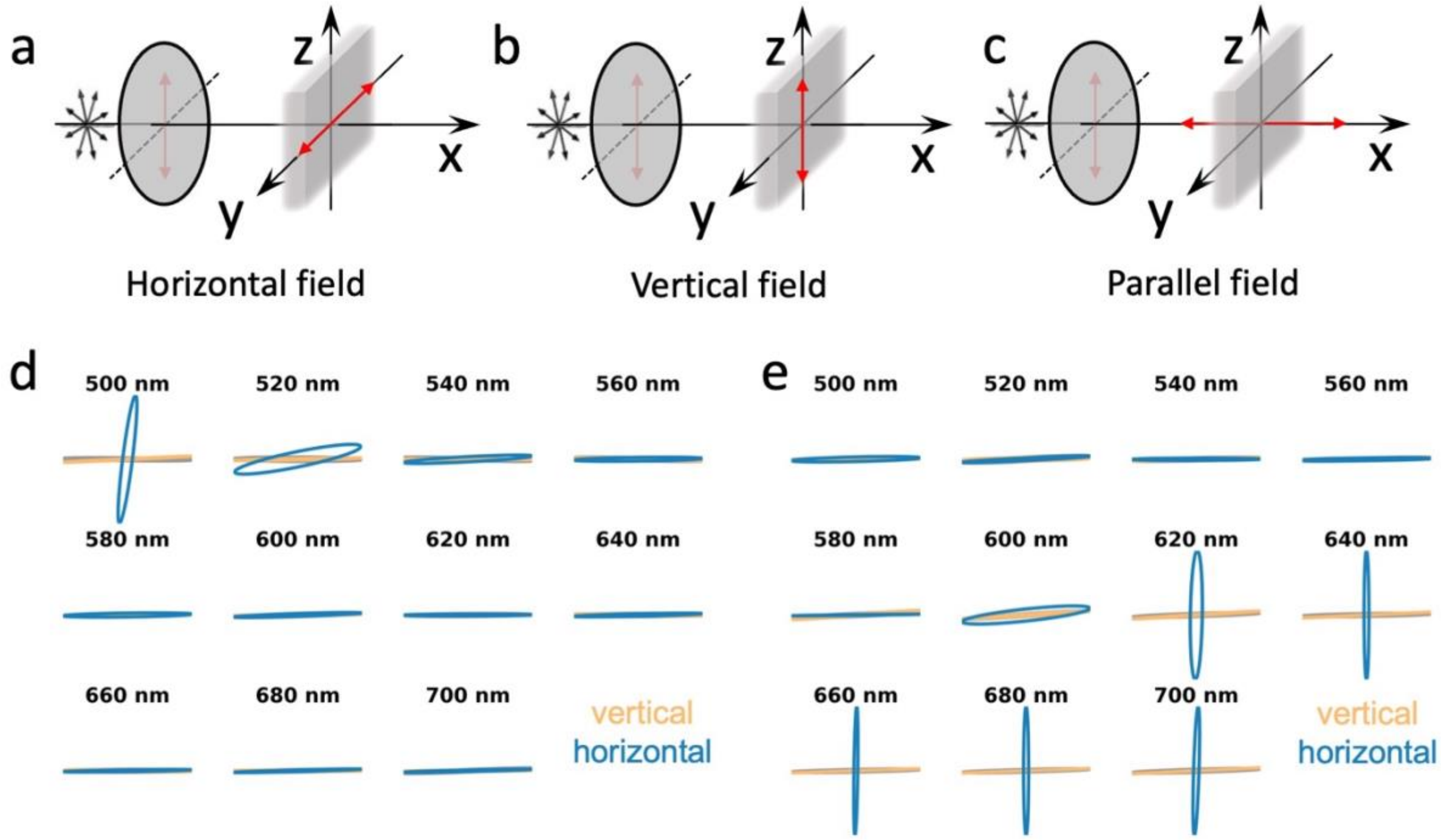

**Figure S3.** Schematic illustration of magnetic field directions in measuring the polarization states of light: **(a)** horizontal, **(b)** vertical, **(c)** parallel magnetic field. Polarization ellipses of nematic superstructures composed of **(d)** magnetic and **(e)** hybrid nanorods. Vertical (yellow color) and horizontal (blue color) magnetic fields were used to align the nanorods. The wavelength of incident light is given in each panel. The incident light was horizontally polarized. The DOP changes of the incident light after transmission through the plasmonic nematic superstructures are shown in Figures 2d and 2e.

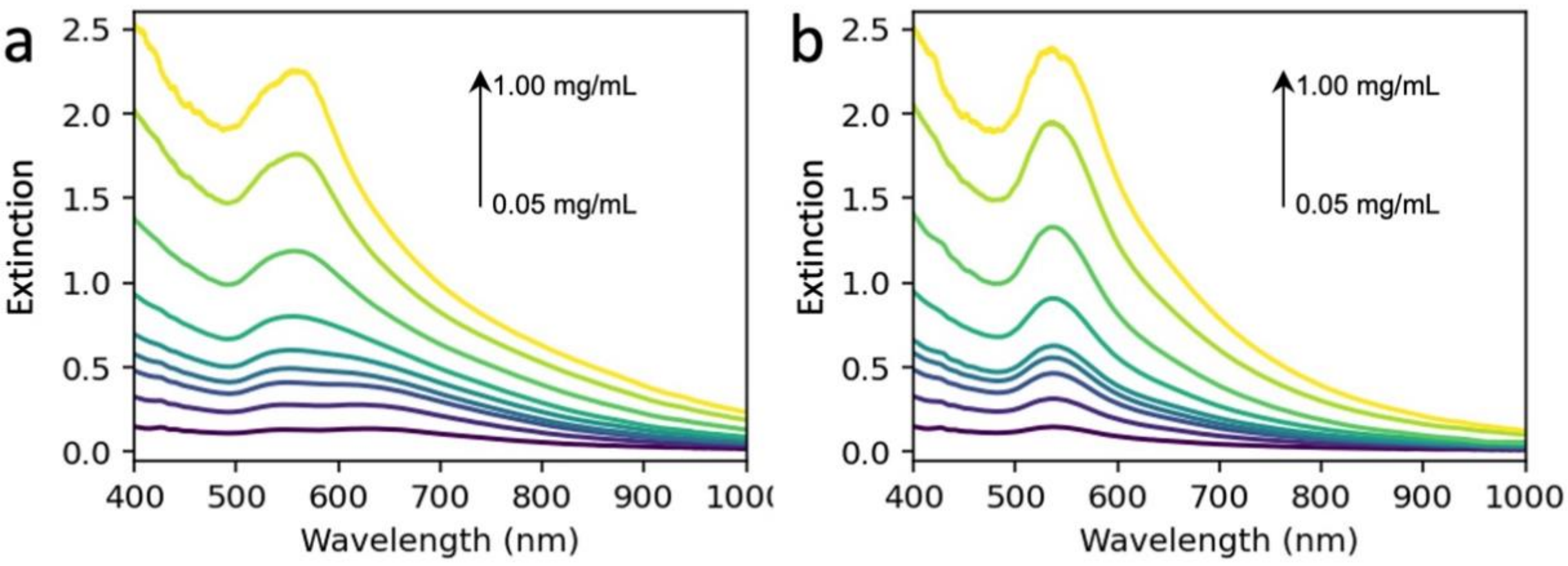

**Figure S4.** Extinction spectra of hybrid nanorods under a magnetic field that is (a) perpendicular and (b) parallel to the direction of the incident light. The concentration of the spectra increases from 0.05 mg/mL to 0.13 mg/mL, 0.17 mg/mL, 0.20 mg/mL, 0.25 mg/mL, 0.33mg/mL, 0.50 mg/mL, 0.67 mg/mL, and 1.00 mg/mL.

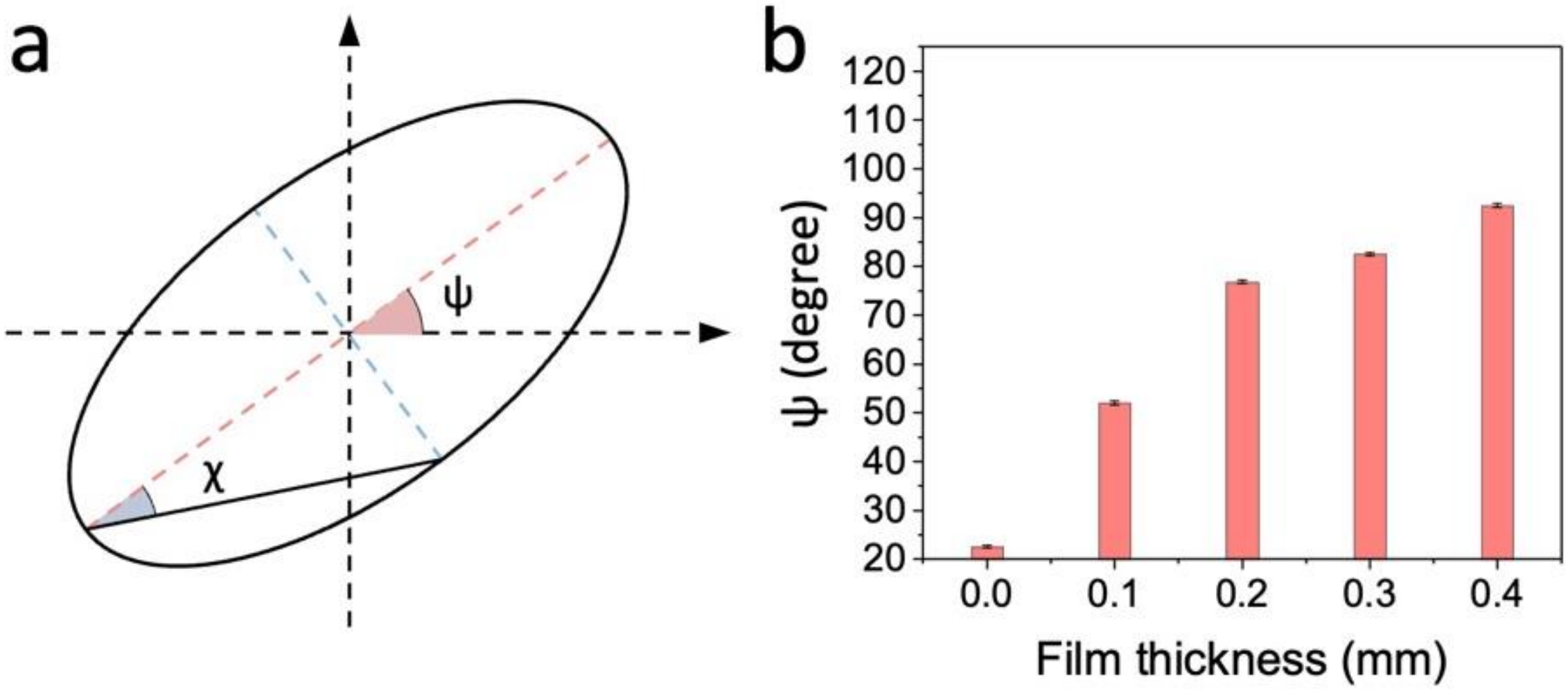

**Figure S5. (a)** Schematic illustration of the polarization ellipse representation for defining the polarization state of light. **(b)** Dependence of the azimuthal angle of the transmitted light on the film thickness

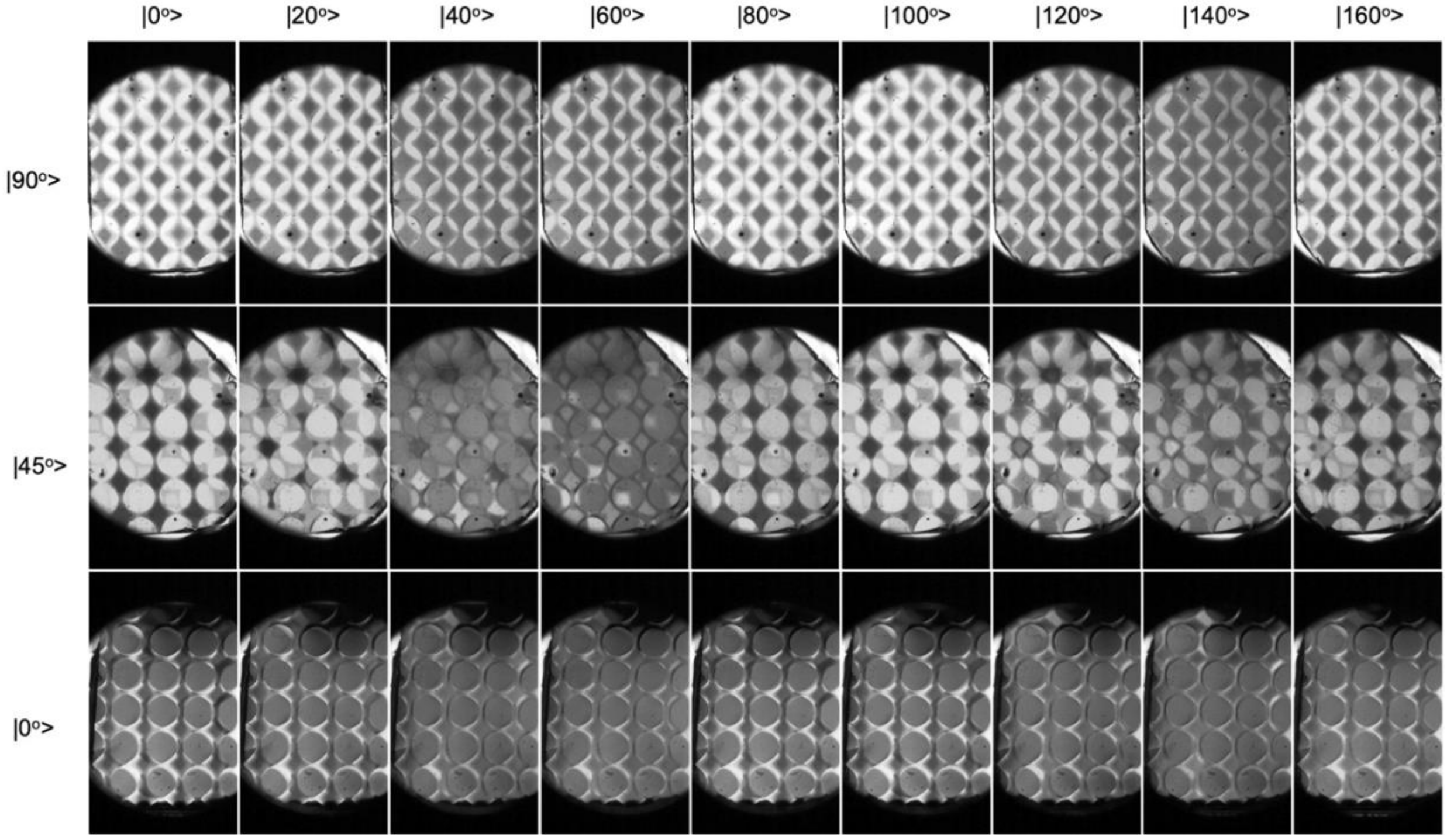

**Figure S6.** Optical images of the DOP for films with different twist angles with illumination at $\lambda_L$. The twist angle is indicated for each row, and the rotation angle of the quarter-wave plate is indicated for each column.

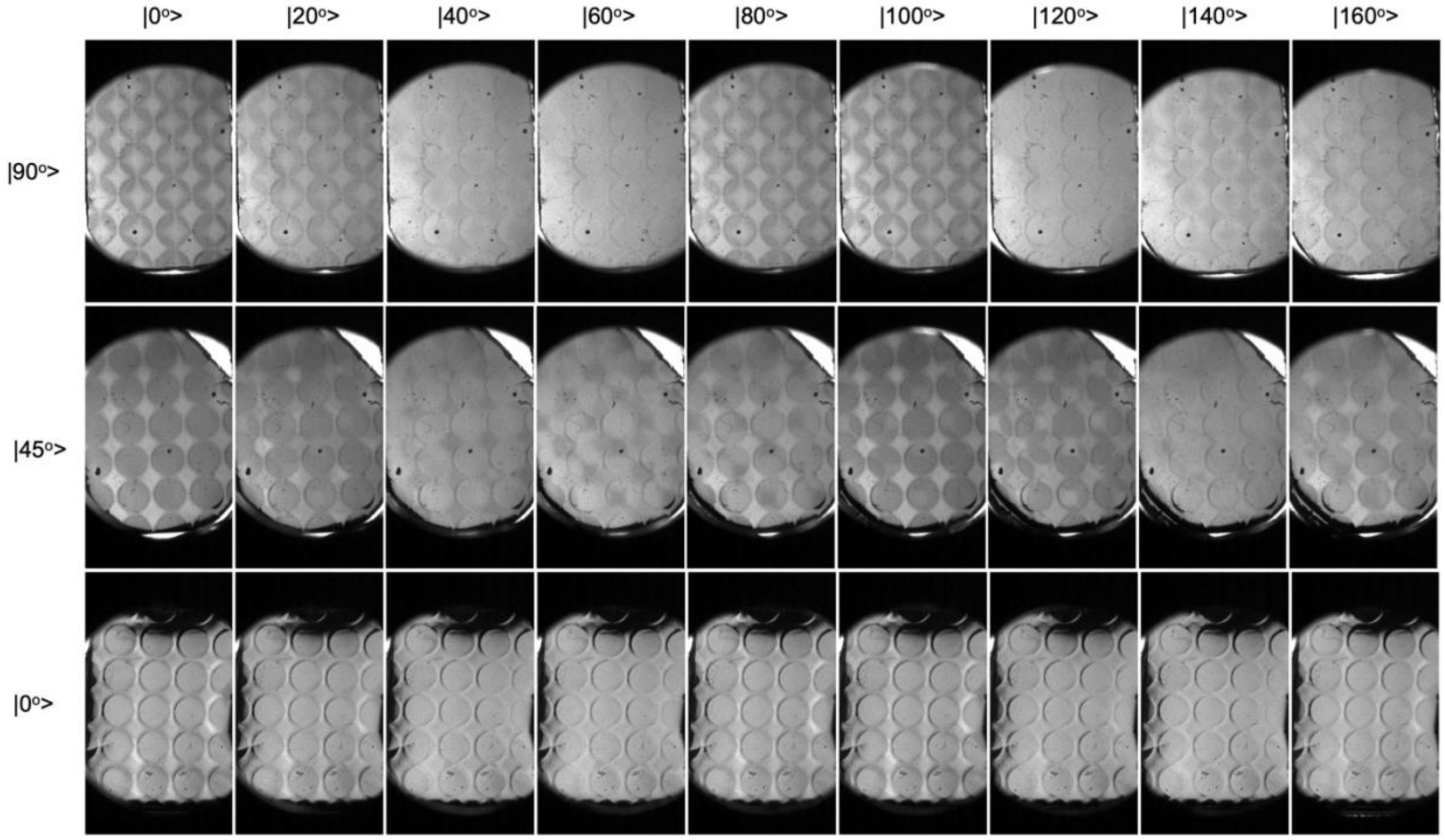

**Figure S7.** Optical images of the DOP for films with different twist angles with illumination at $\lambda_T$. The twist angle is indicated for each row, and the rotation angle of the quarter-wave plate is indicated for each column.

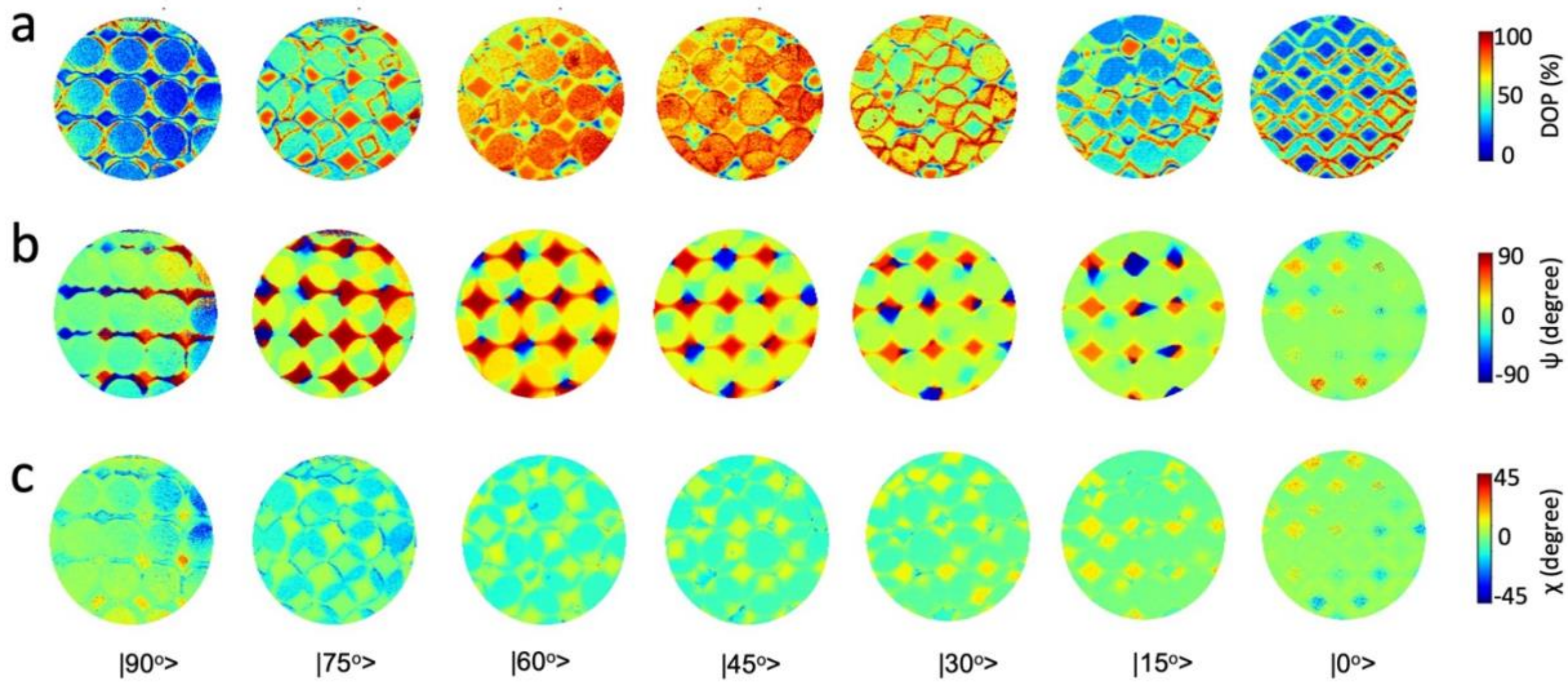

**Figure S8.** Mapping of polarization state of twisted nematic plasmonic superlattices at an illumination wavelength of 685 nm, for (a) the degree of polarization, (b) the orientation angle (ψ), and (c) the ellipticity angle (χ). The twist angle is indicated at the bottom of each column.

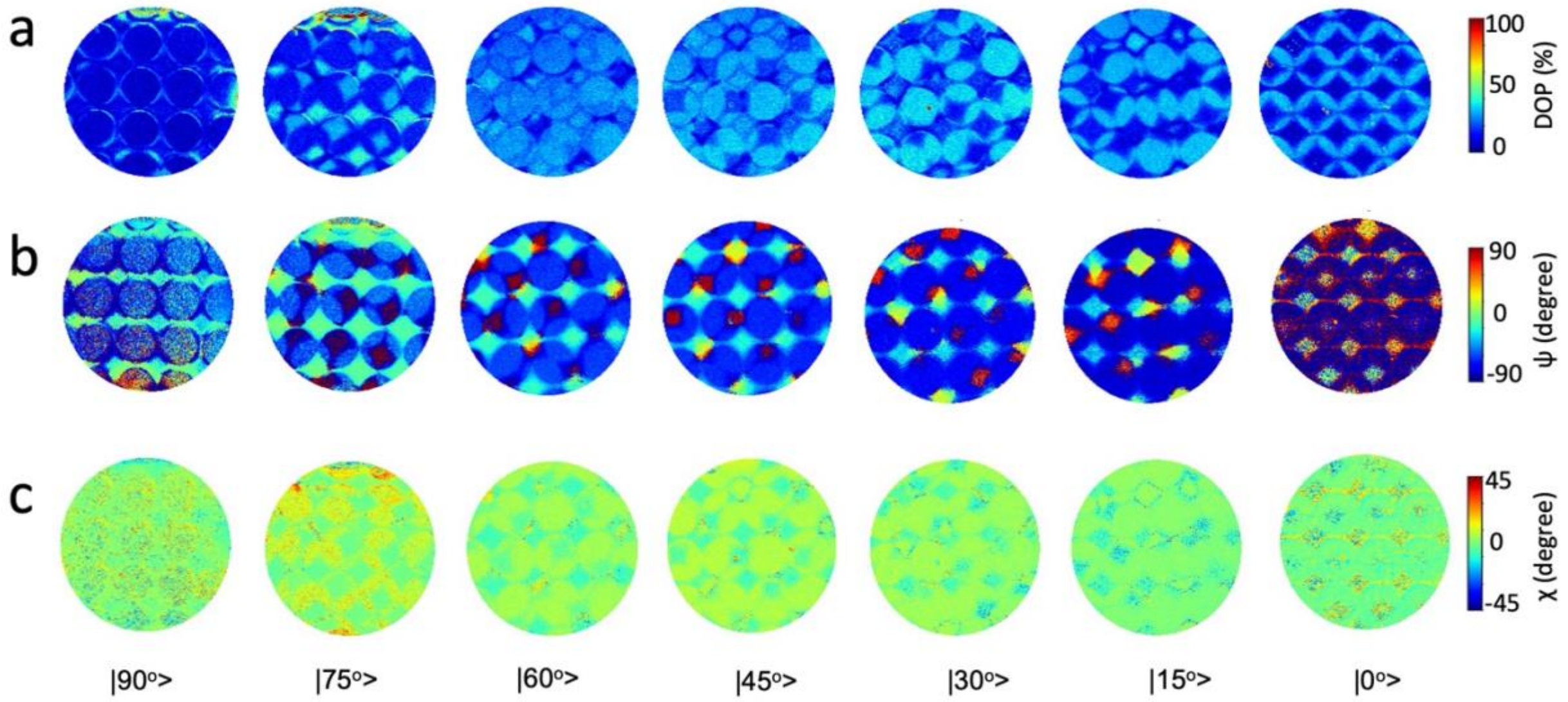

**Figure S9.** Mapping of polarization states of twisted nematic plasmonic superlattices for illumination at 550 nm, for (a) the degree of polarization, (b) the orientation angle (ψ), and (c) the ellipticity angle (χ).

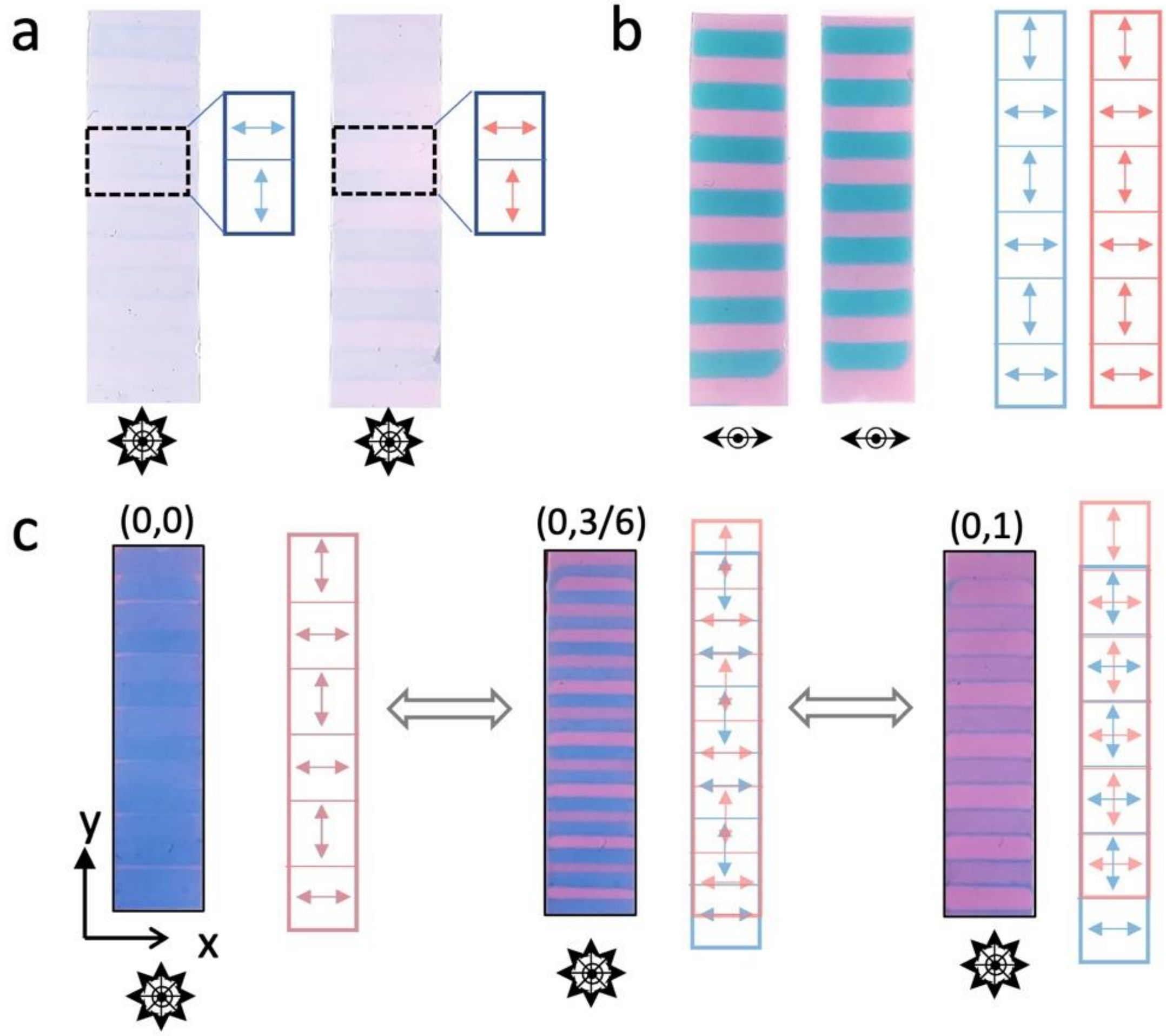

**Figure S10. (a)** Optical images and arrows indicating the nanorod alignment in the two plasmonic films. **(b)** Optical images of the two plasmonic films under polarized light. The nanorod alignment in the film is orthogonal in neighboring color domains. **(c)** Optical images of the twisted bilayer under a translational operation and the corresponding changes in the relative rod alignment in the two films. One translation unit is the width of a single stripe.

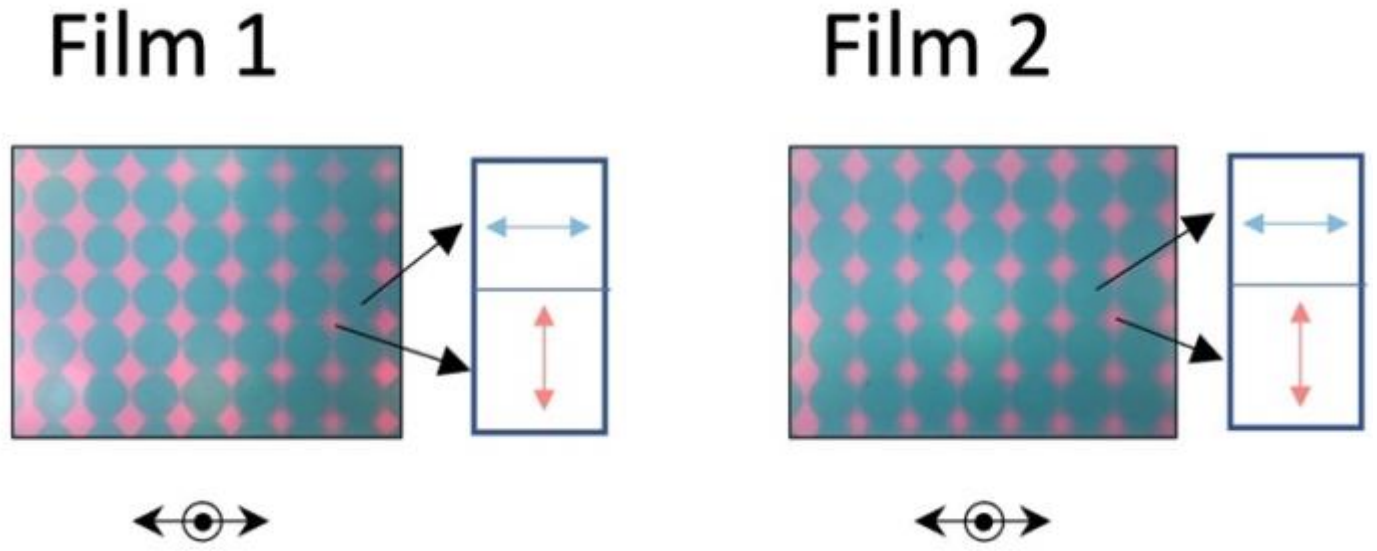

**Figure s11.** Optical images of two plasmonic films with 2D circle lattices. The nanorod alignments in the blue and red areas are illustrated by the arrows.

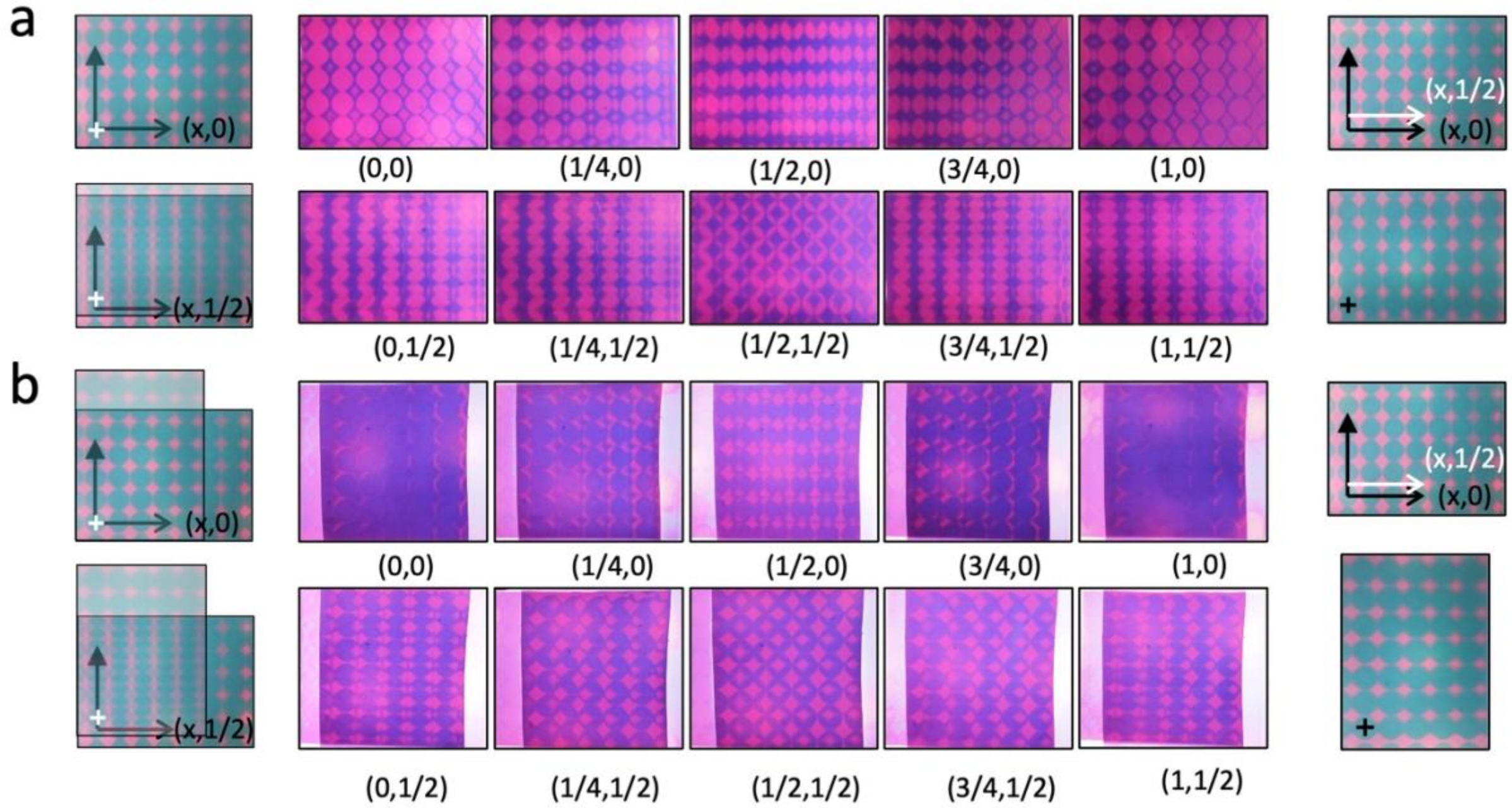

**Figure s12. Highly reconfigurable moiré superlattices in twisted plasmonic bilayers.** Optical images of the moiré superlattices formed by a translational operation under (**a**) parallel and (**b**) perpendicular stacking. The initial stacking position and the final stacking position are illustrated in the left and right panels, respectively. The middle panels show the observed plasmonic moiré patterns under different displacements. The displacement unit is the distance between the circles.

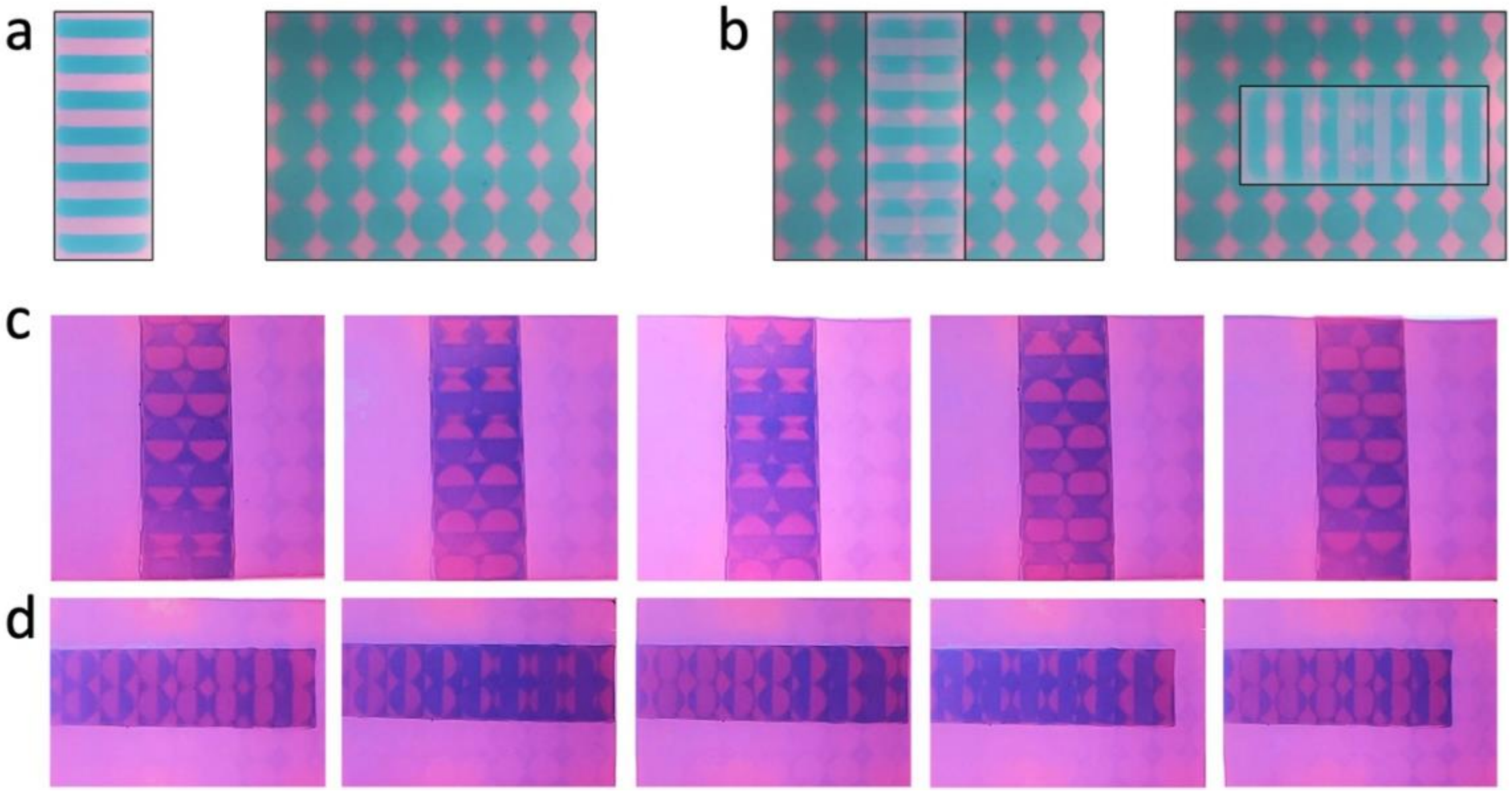

**Figure s13. Moiré superlattices in the twisted plasmonic bilayer that have patterns with different dimensions and symmetries. (a)** Optical images of the two plasmonic films. **(b)** Illustration of the two stacking methods used in preparing these moiré superlattices. The moiré superlattices corresponding to the left and right panel in **(b)** are shown in **(c)** and **(d)**, respectively.